\documentclass[aps,twocolumn,pre,floatfix]{revtex4-2}

\usepackage{xcolor,hyperref}
\hypersetup{
   colorlinks,
   linkcolor={blue!50!black},%{red!80!black},
   citecolor={blue!50!black},
   urlcolor={blue!80!black}
}
\usepackage{epsfig, amsmath}
\usepackage{bm}
\usepackage{soul,ulem}
\usepackage{hyperref}
\usepackage{footmisc}
\usepackage{wrapfig}
\usepackage{amssymb}
\usepackage{babel}
\usepackage{MnSymbol}

\makeatletter
\let\ORIbbl@fixname\bbl@fixname
\def\bbl@fixname#1{%
  \@ifundefined{languagealias@\expandafter\string#1}
    {\ORIbbl@fixname#1}
    {\edef\languagename{\@nameuse{languagealias@#1}}}%
}
\newcommand{\definelanguagealias}[2]{%
  \@namedef{languagealias@#1}{#2}%
}
\makeatother
\definelanguagealias{en}{english}
\DeclareMathAlphabet{\mathitbf}{OML}{cmm}{b}{it}

\DeclareMathOperator{\sgn}{sgn}

\DeclareMathOperator{\arccosh}{arccosh}
\DeclareMathOperator{\arcsinh}{arcsinh}

\DeclareMathOperator{\erf}{erf}
\DeclareMathOperator{\Tr}{Tr}
\DeclareMathOperator{\Supp}{Supp}

\definecolor{darkGreen}{RGB}{0,100,0}

\begin{document}

\title{From non-universal spectra at the RS–RSB transition to the universal \(\omega^4\) regime inside the glassy phase}

\author{Tommaso Pettinari$^{1,2}$}
\email{t.pettinari@uva.nl}
\author{Edan Lerner$^{1}$}
\email{e.lerner@uva.nl}
\author{Gustavo D\"uring$^{3}$}
\email{gduring@uc.cl}
\affiliation{$^{1}$Institute of Theoretical Physics, University of Amsterdam, Science Park 904, 1098 XH Amsterdam, the Netherlands\\
$^{2}$Van der Waals-Zeeman Institute, University of Amsterdam,
Science Park 904, 1098 XH Amsterdam, the Netherlands\\
$^{3}$Instituto de F\'isica, Pontificia Universidad Cat\'olica de Chile, 8331150 Santiago, Chile}

\begin{abstract}
The universal low-frequency nonphononic density of states $\mathcal{D}_{\rm G}(\omega)\!\sim\!\omega^4$ is a robust feature of structural glasses, yet its origin remains poorly understood. Here we investigate this question using the mean-field KHGPS model of interacting anharmonic oscillators. We show that the universal quartic spectrum is deeply rooted in the glassy, replica-symmetry-broken (RSB) phase, whereas the gapless density of states found at the boundary of the replica-symmetric (RS) phase is non-universal and depends sensitively on the details of the quenched disorder. We formulate a simplified variant of the KHGPS model that allows us to characterize the crossover between these two regimes and provides a physical interpretation of the associated crossover frequency scale. We further show that the conventional, \textit{spin-glass-like}, continuous RS--RSB transition, accompanied by diverging susceptibilities, can be suppressed by an appropriate choice of the quenched disorder, allowing for a discontinuous RS--RSB transition even in the absence of an external field. Our results establish a direct connection between the emergence of the universal $\omega^4$ spectrum and the glassy phase, and provide a simple framework for exploring the low-temperature properties of glasses and the soft localized excitations that emerge from bistable local potentials.
\end{abstract}

\maketitle

\section{Introduction}

The thermodynamic and mechanical properties of amorphous solids differ markedly from those of their crystalline counterparts \cite{zeller_thermal_1971, lubchenko_theory_2007, wyart_rigidity_2005}. For example, experiments show that below $1$~K the specific heat and the thermal conductivity of structural glasses \textemdash~amorphous solids obtained by rapidly quenching a supercooled liquid~\textemdash~scale, respectively, linearly and quadratically with temperature, whereas in crystals both quantities scale as $T^3$. The difference between disordered and crystalline solids is so profound that even the origin of their rigidity has to be sought in distinct mechanisms: crystals are rigid due to the breaking of translational symmetry; for glasses, instead, the culprit is the tremendous increase of viscosity during the quench of a liquid, which inhibits relaxation processes, essentially freezing the material in the ``parent'' liquid configuration.

The origin of the anomalous properties of structural glasses with respect to crystalline solids lies in the low-energy spectrum of vibrational excitations, which controls these phenomena. The vibrational spectrum of crystalline solids is well described by Debye's theory \cite{kittel_introduction_nodate}, which accounts for the statistical physics of long-wavelength phonons. It is however evident from numerical simulations \cite{laird_localized_1991, schober_low-frequency_1996, allen_diffusons_1999, tanguy_continuum_2002, silbert_vibrations_2005, shintani_universal_2008, mizuno_continuum_2017, lerner_boson-peak_2023, moriel_boson_2024} that the density of vibrational excitations in glasses, $\mathcal{D}(\omega)$, displays an excess of low-energy modes with respect to the Debye spectrum, $\mathcal{D}_{\rm D}(\omega)\!\sim\!\omega^{d-1}$. This excess was coined the boson peak~\cite{buchenau_neutron_1984, malinovsky_nature_1986} due to the characteristic shape that appears at low frequencies in the reduced density of states $\mathcal{D}(\omega)/\omega^2$. Moreover, through numerical simulations it was established that the additional low-energy excitations differ from Goldstone modes both in their structure, featuring a small disordered core decorated by an algebraically decaying far field (hence the name quasilocalized modes, QLMs), and in their statistics, following a gapless quartic distribution, $\mathcal{D}_{\rm G}(\omega)\!\sim\!\omega^4$ (see Ref.~\cite{lerner_low-energy_2021} for a comprehensive review). This distinctive quartic law has been shown to be a universal property of glass formers, being independent of the microscopic details of the model \cite{lerner_statistics_2016, richard_universality_2020}, of the glass preparation history~\cite{wang_low-frequency_2019,rainone_pinching_2020}, of the presence of long-range order~\cite{lerner_disordered_2022}, and even of the spatial dimension \cite{kapteijns_universal_2018}. 

Many simple models have been proposed in the last few decades to explain the emergence of these excess nonphononic soft modes in glasses, the origin of the boson peak, and, consequently, the thermodynamic anomalies in amorphous solids \cite{buchenau_anharmonic_1991, buchenau_interaction_1992, ramos_density_1993, parshin_soft_1993, kuhn_random_1997, gurevich_anharmonicity_2003,gurarie_bosonic_2003, degiuli_effects_2014, franz_universal_2015, sharma_metastable_2016, baggioli_universal_2019}. Some of these models are able to qualitatively reproduce the glassy nonphononic spectra, while others are effective models that can replicate the desired observables without clarifying completely their microscopic origin. Despite not being fully satisfactory, these efforts greatly contributed to shedding light on the physical mechanisms at play in disordered systems. A notable example is the work by Gurevich, Parshin, and Schober (GPS) who, based on the Soft Potential Model framework~\cite{buchenau_interaction_1992, ramos_density_1993, parshin_soft_1993}, devised a three-dimensional lattice model~\cite{gurevich_anharmonicity_2003} in which a collection of anharmonic oscillators interact through the elastic medium in which they are embedded. Their phenomenological theory was able to correctly reproduce the main features of $\mathcal{D}_{\rm G}(\omega)$ and predict the presence of tunneling two-level systems, which are believed to be the origin of the anomalous linear scaling of the specific heat with temperature universally observed in structural glasses~\cite{parshin_vibrational_2007}.

Another significant step towards a complete understanding of low-energy excitations in structural glasses arrived with the study of fully-connected mean-field models of disordered systems \cite{franz_universal_2015, sharma_metastable_2016}. This class of mean-field models relates the excess of soft modes to marginal stability using the framework of replicas \cite{parisi_theory_2020}, a construction which had already proven itself successful in solving the spin glass problem and that contributed greatly to building the large body of work aiming at describing the phenomenology of the dynamical and thermodynamic glass transition that goes under the name of random first-order transition (RFOT) theory \cite{kirkpatrick_p-spin-interaction_1987, kirkpatrick_comparison_1988, kirkpatrick_scaling_1989, lubchenko_theory_2007, berthier_theoretical_2011}. These mean-field models, though, display a Gardner transition towards a full replica symmetry breaking (fullRSB) phase with diverging susceptibilities (typical of spin glasses, and not found in molecular glasses) and a quadratic density of states, $\mathcal{D}(\omega)\!\sim\!\omega^2$, that features nonphononic modes with a delocalized nature, being at odds with results from numerical simulations and experiments. 

Recently, this picture was challenged by the introduction of a new mean-field model for disordered systems \cite{rainone_mean-field_2021, bouchbinder_low-frequency_2021, folena_marginal_2022}. This model represents a collection of anharmonic soft spins interacting with random couplings and subject to an external field. The model can be seen as an infinite-dimensional version of the GPS model \cite{gurevich_anharmonicity_2003} or as a generalization of the model devised by K{\"u}hn and Horstmann \cite{kuhn_random_1997}, and was hence termed the KHGPS model. It was shown \cite{bouchbinder_low-frequency_2021, folena_marginal_2022} that the model undergoes a zero-temperature transition to a fullRSB phase. Moreover, in a region of the parameter space this transition features a finite spin-glass susceptibility and shows a gapless quartic spectrum, $\mathcal{D}(\omega)\!\sim\!\omega^4$, with localized excitations \textemdash~note that for another range of parameters, the model shows a standard spin-glass transition with diverging susceptibility and a quadratic density of states populated by delocalized modes. The model also displays glassy phenomenology e.g.~when subject to oscillatory shear \cite{suda_yielding_2025} and, when quantized, it can host two-level systems and, consequently, it could account for the anomalous linear dependence of the specific heat on temperature \cite{maimbourg_two-level_2024}. For all these reasons, the KHGPS model emerges as a strong candidate to unify a broad range of disordered systems and explain the glassy phenomenology. Notably, first steps towards a renormalization-group treatment of this model have been taken, breathing new life into research on the spin-glass transition in a field in finite dimensions \cite{urbani_field_2022}. 

In the original work of Ref.~\cite{bouchbinder_low-frequency_2021}, the replica-symmetric (RS) phase was studied, revealing a gapless quartic spectrum, $\mathcal{D}(\omega)\!\sim\!\omega^4$, at the boundary of the replica-symmetry-broken (RSB) phase. However, as we show in this work, this quartic spectrum is not universal; instead, it depends sensitively on the precise form of the  distribution $P(\kappa)$ of local stiffnesses $\kappa$, i.e.~on the choice of quenched disorder of this model. Here we study the replica-symmetric phase and its boundary for a general stiffness distribution $P(\kappa)$. We show that, by varying $P(\kappa)$, one can not only modify the quartic spectrum $\mathcal{D}(\omega)\!\sim\!\omega^4$, but also progressively shrink the region in which the standard spin-glass transition takes place---characterized by a diverging susceptibility and a gapless quadratic spectrum, $\mathcal{D}(\omega)\!\sim\!\omega^2$---until this region disappears completely.

Interestingly, we further find that deep inside the RSB phase, the quartic spectrum is recovered in the KHGPS model, independently of the specific choice of stiffness distribution. Unfortunately, a complete solution of the RSB phase is still lacking. However, inspired by the regime in which the diagonal terms of the Hessian matrix dominate the spectrum, we introduce a simplified model showing that a finite density of oscillators at vanishing effective stiffness is responsible for the universal gapless quartic spectrum. Moreover, this simplified model allows us to determine the crossover scale from the non-universal gapless distribution, similar to that observed at the boundary of the replica-symmetric phase in the KHGPS model, to the universal quartic regime. We then test this prediction in the full KHGPS model and find good agreement. This crossover supports the possibility that real-world and laboratory glasses reside deep within a marginal phase characterized by full replica-symmetry breaking.

The structure of this paper is as follows: in Sec.~\ref{original_KHGPS} we introduce the KHGPS model and summarize previous results of the replica calculation in the RS phase, extending them to a general stiffness distribution $P(\kappa)$. We explicitly show how the spectrum of excitations depends on the details of the distribution of oscillators' stiffness as well as how the phase diagram of the model is modified. Our theoretical results are then validated using numerical simulations. In Sec.~\ref{simplified_khgps}, we introduce a variant of the KHGPS model, referred to in what follows as the ``simplified KHGPS'' model. We study the model analytically, in particular in the zero-temperature limit, obtaining its spectrum of excitations and we explicitly show how this spectrum depends on the details of the distribution $P(\kappa)$ of oscillators' stiffnesses $\kappa$. In Sec.~\ref{crossover} we illustrate how the crossover to the universal quartic scaling $\mathcal{D}(\omega)\!\sim\!\omega^4$ takes place when bistabilities arise in the effective local potential. We finally illustrate how the same mechanism that produces the crossover from a distribution-dependent density of states to the universal quartic scaling applies also in the case of the KHGPS model. We conclude with a discussion on the obtained results in Sec.~\ref{discussion}.

\section{The KHGPS model}\label{original_KHGPS}

We start by providing a brief introduction of the KHGPS model, and review some previous results regarding its properties. The model consists of a collection of soft spins, or anharmonic oscillators, characterized by disordered pairwise interactions. The KHGPS model is defined by the following Hamiltonian
\begin{equation}\label{eq:KHGPS_hamiltonian}
    \mathcal{H} = \sum_i \left( \frac{1}{4!} x_i^4 + \frac{\kappa_i}{2} x_i^2 - h x_i \right) + \sum_{i < j} J_{ij} x_i x_j\,,
\end{equation}
in which the coupling terms $J_{ij}$ follow a Gaussian distribution with zero mean and variance $J^2/N$, and $h$ represents a constant external field. We consider the stiffnesses $\kappa_i$ to be drawn from an arbitrary power-law distribution,
\begin{equation}\label{eq:stiffness_distribution}
    P(\kappa) = \frac{\alpha + 1}{(\kappa_M - \kappa_m)^{\alpha + 1}} (\kappa - \kappa_m)^\alpha \,,
\end{equation} 
with $\kappa_m$ and $\kappa_M$ respectively representing the lower and upper bound of the support of $P(\kappa)$. From the point of view of structural glasses, the disordered interactions and the field in which the oscillators are embedded represent the frustration that builds up during the quench of a supercooled liquid, modifying the spectrum of excitations of the system with respect to the parent distribution $P(\kappa)$. Previous analyses of the model \cite{bouchbinder_low-frequency_2021, rainone_mean-field_2021, folena_marginal_2022} focused on the uniform case, $\alpha\!=\!0$. In our current study, instead, we extend our attention to other distributions, in order to see how they affect the spectrum of low-energy excitations of the system. As in previous studies, we fix the prefactor of the anharmonic term $x^4/4!$ to unity.

We are interested in the distribution of eigenvalues of the Hessian matrix 
\begin{equation}
    \mathcal{M}_{ij} = J_{ij} + \delta_{ij}(\kappa_i + \frac{1}{2}x_i^2)
\end{equation}
in the zero-temperature limit. The Hessian is the sum of a diagonal matrix $\mathcal{A}_{ij}$, with diagonal elements $a_i\!=\!\kappa_i\!+\!\frac{1}{2} x_i^2$, and a matrix $J_{ij}$ from the Gaussian Orthogonal Ensemble (GOE). Following Ref.~\cite{bouchbinder_low-frequency_2021}, it is possible to study the spectrum of the Hessian $\mathcal{M}$ through the resolvent technique \cite{bun_cleaning_2017,bouchbinder_low-frequency_2021}. Two different regimes were found: (i) in one region of the $(J,h)$ parameter space, the GOE component $J_{ij}$ of the Hessian dominates the lower edge of the spectrum, which displays the square-root behavior of the Wigner semicircle law, $\rho_\mathcal{M}(\lambda)\!\sim\!(\lambda\!-\!\lambda_m)^{1/2}$ \cite{pastur_eigenvalue_2011} (with $\lambda_m$ the lower bound of the support). (ii) In the rest of the parameter space, it is the contribution from the diagonal matrix $\mathcal{A}_{ij}$ that dominates the lower edge of the spectrum, which is consequently determined by the distribution of diagonal elements $P(a)$.  We dub the two aforementioned regimes GOE-like and DIAG-like, respectively. These two regimes are separated by a line in the $(J,h)$ plane, whose shape is determined by the self-consistent equation
\begin{equation}\label{eq:Lambdazero}
    \Lambda(J,h) = 1 - J^2\int_{a_m}^{a_M} da \, P(a) \frac{1}{(a - a_m)^2} = 0 \quad ,
\end{equation}
with $a_m$ the lower bound of the support of $P(a)$ (see Appendix \ref{app:spectrum_KHGPS} for detailed calculations, as well as the original article \cite{bouchbinder_low-frequency_2021}, in particular Appendix A therein). For $\Lambda\!<\!0$ the system sits in the GOE-like regime, while for $\Lambda\!>\!0$ the spectrum is DIAG-like. This general picture holds for any arbitrary stiffness distribution and the functional form of $P(\kappa)$ only affects the extension of the two regions of the parameter space. As we will show in the following sections, limiting cases where one of the two regimes takes over the whole parameter space can also be obtained by specific choices of the distribution (for example, the original KH model \cite{kuhn_random_1997}, in which $\kappa_m\!=\!\kappa_M\!=\!1$, always sits in the GOE-like regime, as explained in Ref.~\cite{bouchbinder_low-frequency_2021}). The physical properties in these two regimes are very different, and we will analyze them in more detail in what follows.

\subsection{A brief review of the replica-symmetric phase and its breaking}\label{RS_and_its_breaking}

To better visualize the mechanisms at play in the two different regimes, it is possible to recast the system in a simplified form. Using the replica framework, it is possible to show (see Appendix \ref{app:RS_RSB} for more details, or Ref.~\cite{bouchbinder_low-frequency_2021}, Appendix B therein) that the KHGPS model is equivalent to a collection of independent oscillators subject to the effective potential
\begin{equation}\label{eq:eff_pot_KHGPS}
    V_{\rm eff} (x) = \frac{1}{4!}x^4 + \frac{m}{2}x^2 - H x
\end{equation}
with
\begin{align}
    m =~& \kappa - J^2 \chi \qquad ,
    \label{eq:eff_pot_KHGPS_m}\\
    H =~& h + J z\sqrt{\tilde{q}} \qquad . \label{eq:eff_pot_KHGPS_H}
\end{align}
where $z$ is a random variable normally distributed around zero with unit variance and the newly introduced parameters $\tilde{q}$ and $\chi$ originate from the correlation between replicas. The effective stiffness is shifted by the disorder-induced amount $J^2 \chi$, hence its distribution is
\begin{equation}
    P_m(m) = P(m + J^2\chi) \quad ,
\end{equation}
while the field has a Gaussian distribution, $P_H(H)\!=\!\mathcal{N}(h, J^2 \tilde{q})$. When $m_m\!\ge\!0$, the parameters $\tilde{q}$ and $\chi$ are defined by the following self-consistent equations
\begin{eqnarray}
    &\tilde{q} = \big\langle [x^*(z,m)]^2\big\rangle_{z,m} \quad , \label{eq:self_cons_q}\\
    &\chi = \big\langle \frac{1}{V''_{\rm eff}[x^*(z,m)]}\big\rangle_{z,m} \quad , \label{eq:self_cons_chi}
\end{eqnarray}
where $x^*(z,m)$ is the absolute minimum of the potential, given the effective stiffness $m$ and the additional effective field $z$. We can interpret $\tilde{q}$ as the mean squared displacement (averaged over the disorder) of the effective oscillators, and $\chi$ as their average susceptibility, with
\begin{equation}
    V''_{\rm eff}|_{x^*} = m + \frac{(x^*)^2}{2} \equiv \tilde{a} \,,
\end{equation}
or, equivalently, using the definition of the diagonal elements of $\mathcal{A}$, $\tilde{a}\!=\!a\!-\!J^2 \chi$.

The solution that we just spelled out is valid only in the replica-symmetric phase, where the potential energy landscape is characterized by a unique ground state. At a sufficiently high, $h$-dependent value of the interaction strength $J$, denoted $J_c(h)$, the solution becomes unstable. The transition to the replica-symmetry-broken (RSB) phase is marked by the replicon mode of the replicated free energy Hessian, $\lambda_R$, becoming negative \cite{nishimori_statistical_2001}, which destabilizes the solution. In our system, the expression for the replicon $\lambda_R$ reads (see Ref.~\cite{bouchbinder_low-frequency_2021} for the calculation details)
\begin{equation}\label{eq:replicon_integ}
    \lambda_R = 1 - J^2\int_{a_m}^{a_M} da \, P(a) \frac{1}{(a - J^2\chi)^2} \quad .
\end{equation}
We note that the replicon $\lambda_R$ differs from the expression of $\Lambda$ in Eq.~\eqref{eq:Lambdazero} only in the denominator, where $a_m$ has been substituted by $J^2\chi$. Since $a\!=\!m\!+\!\frac{1}{2}(x^*)^2\!+\!J^2\chi$, as long as $m_m\!\ge\!0$ ($m_m$ is the lower bound of the support of $P_m$) we have $a_m\!\ge\!J^2\chi$, and consequently
\begin{equation}\label{eq:lambdas_inequality}
    \lambda_R \ge \Lambda \quad ,
\end{equation}
with the equality holding when $m_m$ vanishes.

The breaking of replica symmetry can happen in two different ways, depending on the regime (see again Appendix \ref{app:RS_RSB} for a more detailed derivation): (i) in the GOE-like regime, $\lambda_R$ smoothly crosses over to negative values. At the transition, when $\lambda_R\!=\!0$, the density of states becomes gapless showing a quadratic scaling, $\mathcal{D}(\omega)\!\sim\!\omega^2$. This corresponds to a standard RSB transition for spin glasses (see e.g.~\cite{franz_universal_2015, sharma_metastable_2016}). (ii) In the DIAG-like regime, instead, upon increasing the coupling $J$ we observe that the lower bound of the support of the effective stiffness, $m_m\!=\!\kappa_m\!-\!J^2 \chi$, reaches zero while the replicon $\lambda_R$ is still finite (and positive). Above this value of disorder $m_m$ becomes negative. This produces double wells in the effective potential, and consequently the global minimum jumps discontinuously when the effective field $H$ (cf.~Eq.~(\ref{eq:eff_pot_KHGPS_H})) switches sign. Ultimately, this results in a Dirac delta in the expression for the replicon, Eq.~\eqref{eq:replicon_integ}, which formally diverges to $-\infty$ when $m_m\!<\!0$, breaking the stability of the replica-symmetric solution. Also in this regime we observe that the gap in the density of states closes exactly at the RSB transition (when $m_m\!=\!0$), producing a pseudogap, which in the case analyzed in Ref.~\cite{bouchbinder_low-frequency_2021} (uniform distribution of stiffness $P(\kappa)$) follows a quartic scaling, $\mathcal{D}(\omega)\!\sim\!\omega^4$. 

\begin{figure}
\centering
{\includegraphics[scale=0.45]{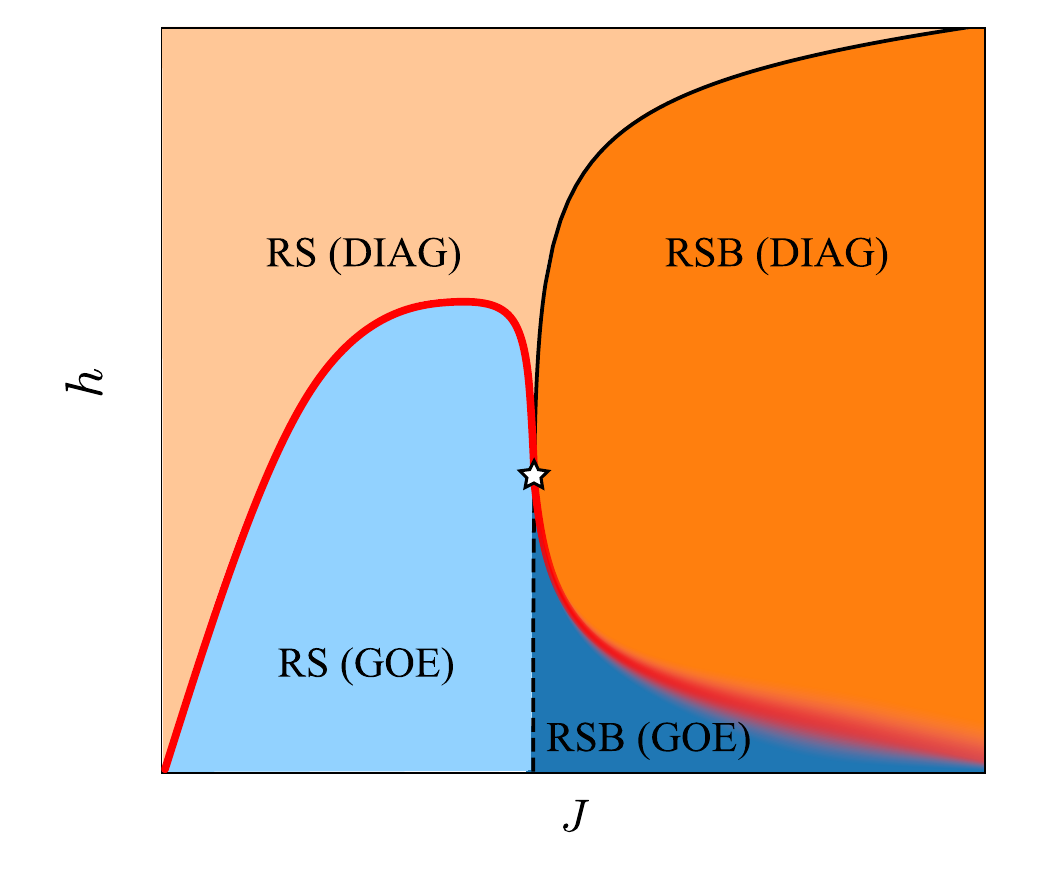}}
\caption{Qualitative sketch of the phase diagram of the KHGPS model, in which we represent the twofold division of the parameter space into DIAG--GOE regimes (respectively in orange and blue, separated by a red line) and into RS--RSB phases (the former colored in a lighter tone, and the latter in a darker one). The shape of the curves separating the various regions changes with the parameters $\alpha$, $\kappa_m$ and $\kappa_M$ of the system, and so does their intersection -- the special point marked by the white star (e.g. the figure represents a sketch of the phase diagram obtained for $\alpha\!=\!0.6$, $\kappa_m\!=\!0.1$ and $\kappa_M\!=\!1.0$). Nonetheless, the RSB transition in the GOE-like regime is always continuous (dashed black line) while in the DIAG-like regime it is always discontinuous (solid black line), in the sense defined in the text. In the RS phase, where the replica ansatz provides us with an expression for $P(a)$, it is possible to obtain the red line designating the DIAG--GOE boundary, given by Eq.~\eqref{eq:Lambdazero}. In the RSB phase the marginal stability \cite{folena_marginal_2022} of the system forces $\lambda_R\!=\!0$, but the RS inequality $\Lambda\!\le\!\lambda_R$ no longer holds: $\lambda_R$ now also contains the contribution of the jumps of the global minimum of double-well effective potentials, which does not enter the Hessian. Hence, in the RSB phase $\Lambda$ becomes positive as soon as double-well effective potentials appear, i.e.\ when $J^2\chi$ reaches $\kappa_m$, and the line separating the GOE-like from the DIAG-like behavior must be found numerically (only qualitatively sketched in the figure).}
\label{fig:phase_diagram_qualitative}
\end{figure}

Given the two different mechanisms through which the RS solution loses stability, we name the two distinct types of transition to the RSB phase as \emph{continuous} for the standard GOE-like regime and \emph{discontinuous} for the DIAG-like regime (the continuity refers to the replicon $\lambda_R$ across the transition). The two portions of the transition line meet at a special point, in which both $m_m\!=\!0$ and $\lambda_R\!=\!0$. A direct consequence of this is that at the special point Eq.~\eqref{eq:Lambdazero} is satisfied and $\Lambda\!=\!0$. Hence, the special point lies on the intersection between the critical line, where the replica symmetry breaks down, and the line separating the GOE-like from the DIAG-like regime (see Fig.~\ref{fig:phase_diagram_qualitative} for a schematic illustration of the phase diagram).

We will now focus our attention on the RS--RSB transition line and in particular on the division in two sub-regions, studying how the location of the special point which separates the two is affected by the form of the stiffness distribution $P(\kappa)$. We leave a detailed study of the evolution of the $\Lambda\!=\!0$ line for future work, since our main focus is on the \emph{discontinuous} RSB transition, for which locating the special point is sufficient.

%%%%%%%%%%%%%%%%%%%%%%%%%%
%%%%%%%%%%%%%%%%%%%%%%%%%%

\subsection{The phase diagram}\label{phase_diag}

We are interested in understanding how changes in the stiffness distribution $P(\kappa)$ affect the boundaries of the RSB phase, and in particular how the special point separating the continuous from the discontinuous transitions --- marked by the white star in Fig.~\ref{fig:phase_diagram_qualitative} --- moves in the parameter space, namely in the $(J,h)$-plane.

\begin{figure*}
\centering
{\includegraphics[scale=0.55]{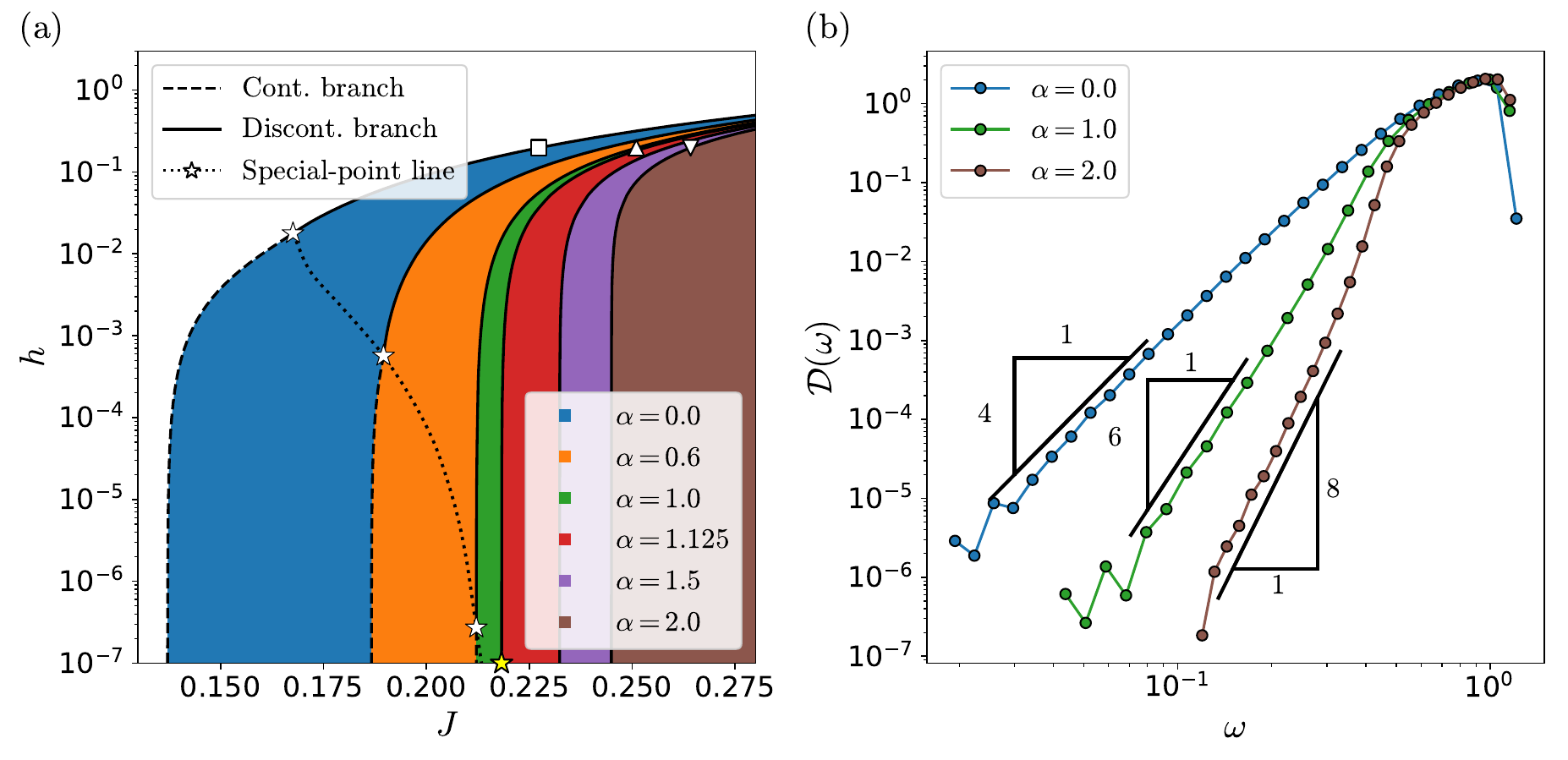}}
\caption{(a) Phase diagram of the KHGPS model (cf.~Eq.~(\ref{eq:KHGPS_hamiltonian})) in the $(J,h)$ plane for $\kappa_m\!=\!0.1$, $\kappa_M\!=\!1$ and for different values of the exponent $\alpha$, ranging from $\alpha\!=\!0$ (i.e. a uniform distribution of stiffnesses) to $\alpha\!=\!2$. The line $J_c(h)$ separates the convex-landscape RS phase from the rough-landscape RSB phase (which is colored in the picture). Along the dashed branch of the line, $\mathcal{D}(\omega)\!\propto\!\omega^2$, while $\mathcal{D}(\omega)\!\propto\!\omega^{4+2\alpha}$ on the solid branch. The two regions of the transition line are separated by the special point in which both the replicon and the lower bound of the effective stiffness distribution are equal to zero: $\lambda_R\!=\!0$, $m_m\!=\!0$. We denote with a dotted line the location of the special point when $\alpha$ is varied, marking it with white stars $\scriptstyle\largestar$ for the selected values of $\alpha$ whose phase diagram is shown. Note that the line plunges towards $h\!=\!0$ when the exponent grows: in fact, as explained in Sec.~\ref{phase_diag}, the line intersects the zero-field axis for a finite exponent $\alpha_0$, which for our parameter choice corresponds to $\alpha_0\!=\!1.125$ (we represent the $h\!=\!0$ end of the line of special points, not representable on a log-scale, with a yellow star). No continuous RSB transition exists for exponents larger than that (see Appendix \ref{app:RS_RSB} for a 3D visualization of the phase diagram). The square $\square$ and the upper ($\triangle$) and lower ($\bigtriangledown$) triangles mark the parameters $(J,h)$ used for the systems simulated in panel (b) of the figure. (b) Density of states of the KHGPS system for the three sets of parameters marked in panel (a), each with a different value of the exponent $\alpha$, but all of them on the RSB transition line: the square corresponds to $\alpha\!=\!0.0$, $J\!=\!0.2273$ and $h\!=\!0.1975$; the upper triangle to $\alpha\!=\!1.0$, $J\!=\!0.2512$ and $h\!=\!0.1975$; the lower triangle to $\alpha\!=\!2.0$, $J\!=\!0.2642$ and $h\!=\!0.1975$. The plots are obtained by averaging over $50\,000$ samples with $N\!=\!10\,000$ oscillators each. The $\alpha$-dependent scaling of the density of states, $\mathcal{D}(\omega)\!\propto\!\omega^{4+2\alpha}$, is confirmed.}
\label{fig:phase_diagram}
\end{figure*}

We extend the parameter space of the KHGPS model to a third dimension, with the exponent $\alpha$ characterizing $P(\kappa)$ (cf.~Eq.~(\ref{eq:stiffness_distribution})) as the additional control parameter. We plot a projection of the phase diagram on the $(J,h)$ plane for selected values of $\alpha$ in Fig.~\ref{fig:phase_diagram}(a), showing how the RSB phase shifts to higher values of disorder when $\alpha$ increases. The transition lines were found numerically, following the algorithm introduced in Ref.~\cite{bouchbinder_low-frequency_2021}: for a certain value of the exponent $\alpha$, we either fix $J$ or $h$ and then iteratively solve (with a damped fixed-point iteration algorithm) the self-consistent equations \eqref{eq:self_cons_q} and \eqref{eq:self_cons_chi}, together with the replicon Eq.~\eqref{eq:replicon_integ}, under the constraint of $m_m\!=\!0$ for the discontinuous transition line, or of $\lambda_R\!=\!0$ for the continuous one. By varying $\alpha$, we observe that the special point separating the two parts of the transition line is plunging towards smaller and smaller values of the external field $h$ as we increase $\alpha$, tracing a line that intersects the $h\!=\!0$ plane at a finite $\alpha_0$.

Remarkably, we can obtain an analytical expression for the critical value $\alpha_0$. Considering the three equations \eqref{eq:self_cons_q}, \eqref{eq:self_cons_chi} and \eqref{eq:replicon_integ}, together with the three combined constraints $h\!=\!0$, $\lambda_R\!=\!0$ and $m_m\!=\!0$ (this last condition is equivalent to $\kappa_m\!=\!J^2 \chi$) we have that: (i) when the external field is null, the integral of the self-consistent equation for $\tilde{q}$, Eq.~\eqref{eq:self_cons_q}, collapses to zero; (ii) given $\tilde{q}\!=\!0$ the effective potential $V_{\rm eff}$ (cf.~Eq.~\eqref{eq:eff_pot_KHGPS}) becomes symmetric and $x^*(z,m)\!=\!0$; (iii) the integrals in the expressions for $\lambda_R$ and $\chi$, Eqs.~\eqref{eq:replicon_integ} and \eqref{eq:self_cons_chi} respectively, become
\begin{eqnarray}\label{eq:lambda_and_chi_h0}
    \lambda_R = 1 -  J^2 \frac{\alpha_0 + 1}{\alpha_0 - 1} w^{-2} = 0 \\
    J^2\chi  = J^2 \frac{\alpha_0 + 1}{\alpha_0} w^{-1} = \kappa_m 
\end{eqnarray}
where we defined $w\!=\!\kappa_M\!-\!\kappa_m$ (see Appendix \ref{app:RS_RSB}). Combining the two expressions, we obtain that the line of special points touches $h\!=\!0$ when
\begin{equation}
    \alpha_0 = \frac{w}{w-\kappa_m} \quad .
\end{equation}
Hence, provided $w\!>\!\kappa_m$ (needed in order to not break replica symmetry and have $\lambda_R\!>\!0$), we have that for exponents $\alpha\!>\!\alpha_0$, no continuous replica symmetry breaking can take place in the KHGPS model, and that the only viable transition to the RSB phase is the discontinuous one (for the choice of parameters $\kappa_m\!=\!0.1$, $\kappa_M\!=\!1$ used in Fig.~\ref{fig:phase_diagram}, this happens at $\alpha_0\!=\!1.125$).

%%%%%%%%%%%%%%%%%%%%%%%%%
%%%%%%%%%%%%%%%%%%%%%%%%%

\subsection{The density of states in the DIAG-like regime: non-universality at the RSB transition}\label{KHGPS_non_universal}

Let us now focus our attention on the spectrum of excitations of the system, in particular in the regime dominated by the diagonal components of the Hessian matrix. In this region of the parameter space, the system develops a power-law spectrum $\rho_{\mathcal{M}}\!\sim\!(\lambda\!-\!\lambda_m)^\nu$, with an exponent $\nu$ determined by the scaling of the distribution of diagonal elements in the Hessian, $P(a)\!\sim\!(a\!-\!a_m)^\nu$. A previous analysis~\cite{bouchbinder_low-frequency_2021} obtained $\nu\!=\!\frac{3}{2}$ when the stiffness distribution $P(\kappa)$ is uniform, i.e.~for $\alpha\!=\!0$, cf.~Eq.~(\ref{eq:stiffness_distribution}). We are interested in understanding how the exponent $\nu$ changes when $P(\kappa)$ changes.

Let us consider again the power-law distribution introduced in the previous section, namely Eq.~\eqref{eq:stiffness_distribution}. Clearly, modifying $P(\kappa)$ does not affect the density of states in the GOE-like regime, since the latter is controlled by the spectrum of the Gaussian matrix $J_{ij}$. On the other hand, varying the exponent $\alpha$ in $P(\kappa)$ has an impact on the distribution of the diagonal elements $a_i\!=\!\kappa_i\!+\!\frac{1}{2}(x^*_i)^2$ (evaluated in the ground state $\{x_i^*\}$), and consequently it might affect the exponent $\nu$ of $P(a)$. 

Consider the effective potential \eqref{eq:eff_pot_KHGPS} and the curvature $\tilde{a}$ at its minimum $x^*$, namely
\begin{equation}
    \tilde{a} = m + \frac{(x^*)^2}{2}\,,
\end{equation}
which, as we have already seen, can be alternatively written as $\tilde{a}\!=\!a\!-\!J^2\chi$, to highlight that $\tilde{a}$ and the diagonal elements $a$ are related by a simple shift. This implies that if we resolve the distribution $P(\tilde{a})$ of the curvatures $\tilde{a}$, we immediately obtain $P(a)$, and hence the spectrum ${\cal D}(\omega)$ of excitations.

We first comment on the support of $P(\tilde{a})$. In a minimum (i.e.~at $x^*$) the curvature is non-negative so the lower bound of the distribution support, $\tilde{a}_m$, must be $\tilde{a}_m\!\geq\!0$. The only way to have a gapless distribution, $\tilde{a}_{m}\!=\!0$, is to have a purely quartic potential, with both $m\!=\!H\!=\!0$. For this reason if $m_m\!>\!0$ (or $m_M\!<\!0$) then the curvature distribution $P(\tilde{a})$ is gapped. We also notice that in the gapless case
\begin{equation}\label{eq:amin}
    \tilde{a}_m = 0 \quad \implies \quad a_m = J^2 \chi \  .
\end{equation}

To compute $P(\tilde{a})$ we first need to resolve the ground state $x^*$ of the effective potential, Eq.~\eqref{eq:eff_pot_KHGPS}. It is given by the solution of the following depressed cubic equation
\begin{equation}\label{eq:depressed_cubic}
	x^3 + 6 m x - 6 H = 0 \quad ,
\end{equation}
whose discriminant is
\begin{equation}
	\Delta \propto m^3 + \frac{9}{8} H^2 \, .
\end{equation}
Whenever $\Delta\!>\!0$, Eq.~\eqref{eq:depressed_cubic} yields a single solution, while the equation has three real roots when $\Delta\!<\!0$. The presence of multiple minima breaks the replica-symmetry ansatz (note that a negative $\Delta$ implies a negative stiffness $m$) and our solution loses validity. For this reason we focus on the case $m\!>\!0$, for which the discriminant is always positive for any value of the field $H$ (see Appendix \ref{app:curv_distribution} for the detailed calculations, including the region with negative discriminant).

Using the known expressions for the roots of cubic equations, we can obtain an explicit form for $x^*(m,H)$. With $x^*$ in hand, we can finally access the distribution of curvatures $P(\tilde{a})$. Following Ref.~\cite{bouchbinder_low-frequency_2021}, we obtain $P(\tilde{a})$ via the cumulative, defined as
\begin{equation}
    G(\tilde{a}) = \int_{0}^{\tilde{a}} d \tilde{a}' \, P(\tilde{a}') \quad .
\end{equation}
In the limit of small curvature $\tilde{a}$, it is possible to show (see again Appendix \ref{app:curv_distribution}) that 
\begin{equation}\label{eq:Gscaling}
    G (\tilde{a}) \sim \tilde{a}^{5/2} P_H(0)\iint  d \eta \, d \xi \,  \xi^{3/2} (\xi \tilde{a} - m_m)^\alpha \quad ,
\end{equation}
where $ P_H(0)\!=\!\frac{1}{ \sqrt{2 \pi J^2 \tilde{q}}} e^{-\frac{h^2}{2 J^2 \tilde{q}}}$ explicitly depends on the control parameters of the system, namely $(J,h)$.

We can extract the scaling of the curvature distribution from Eq.~\eqref{eq:Gscaling} simply by differentiation with respect to $\tilde{a}$. At the RS--RSB transition, which, in the DIAG-like regime (as we stated in Sec.~\ref{RS_and_its_breaking}), coincides with the closure of the gap in the spectrum and is obtained for $m_m\!=\!0$, this gives
\begin{equation}\label{eq:scaling_curvature}
    P(\tilde{a}) \sim \tilde{a}^{3/2 + \alpha} \quad .
\end{equation}
For a flat stiffness distribution, $\alpha\!=\!0$, we retrieve the result of Ref.~\cite{bouchbinder_low-frequency_2021}: a curvature distribution which follows a fixed power-law with exponent $3/2$. Any other scaling of $P_m$, instead, contributes with an additional $\tilde{a}^\alpha$, producing an exponent $\nu\!=\!\frac{3}{2}\!+\!\alpha$. In the DIAG-like regime this exponent coincides with the scaling exponent of the spectrum of the Hessian, $\rho_\mathcal{M}$. Thus, using the relation $\mathcal{D}(\omega)\!=\!2\omega \rho_\mathcal{M}(\omega^2)$, we obtain that the density of states follows 
\begin{equation}\label{eq:alpha-dependent-vdos}
    \mathcal{D}(\omega)\sim\omega^{4+2\alpha} \quad .
\end{equation}
Interestingly, the original KHGPS model exhibits a non-universal density of states at the discontinuous RSB transition, a detail overlooked in Ref.~\cite{bouchbinder_low-frequency_2021}. The spectrum of excitations depends on the exponent $\alpha$ that characterizes the vanishing of the effective stiffness distribution $P_m(m)$ at its lower edge. In Fig.~\ref{fig:phase_diagram}(b) we confirm our analytical predictions by extracting via numerical simulations the density of states of the KHGPS system exactly at the RS--RSB transition.

For completeness, we mention that the scaling seen in Eq.~\eqref{eq:scaling_curvature} at low $\tilde{a}$ is valid only for finite external field $h$ (due to the breaking down of the approximation that leads to Eq.~\eqref{eq:Gscaling} in the limit of vanishing $h$) and that another scaling of the curvature distribution is observable. To understand this, it is useful to consider the extreme case $h\!=\!0$: on the zero-field axis, the only mechanism that breaks the $\mathbb{Z}_2$ symmetry of the oscillators is missing, so the mean square displacement $\tilde{q}$ collapses to zero and the effective field $H$ also vanishes. This makes it much easier to compute the curvature distribution, which simply becomes the effective stiffness distribution, $P(\tilde{a})\!=\!P_m(\tilde{a})$. The alternative scaling of the curvature distribution at zero field $h$ affects the density of states only in the DIAG-like regime, namely when $\alpha\!>\!\alpha_0$ and $\kappa_M\!>\!2\kappa_m$. For this range of the parameters $\kappa_M$ and $\kappa_m$, the density of low-energy excitations follows the scaling
\begin{equation}
    \mathcal{D}(\omega)\sim\omega^{1+2\alpha} \quad .
\end{equation}
A theory for the crossover between these two $\alpha$-dependent scaling regimes at low $h$ can be derived and can be found in Appendix \ref{app:curv_distribution}.

When $m_m\!<\!0$, the replica-symmetric ansatz breaks down, and the solution discussed above no longer applies. Numerical simulations, however, show that deep inside the RSB phase a universal $\mathcal{D}(\omega)\!\sim\!\omega^4$ spectrum is recovered. It has been shown in Ref.~\cite{folena_marginal_2022} that an effective description in terms of ``reconstructed'' non-interacting oscillators remains valid in the RSB phase: the effective stiffness is still $m = \kappa - J^2 \chi$, so that $P_m(m)=P(m+J^2 \chi)$, with $\chi$ now given by the RSB solution, whereas the distribution of the effective fields becomes considerably more involved and depends on $m$. However, a complete theoretical derivation of the universal $\mathcal{D}(\omega)\!\sim\!\omega^4$ spectrum is still lacking.

To make progress, drawing inspiration from the effective potential in Eq.~(\ref{eq:eff_pot_KHGPS}) and from the soft-potential model~\cite{gurevich_anharmonicity_2003}, we introduce a simplified version of the KHGPS model. This simplified model allows us to study the crossover from the non-universal, gapless spectrum observed at the boundary of the RS phase, to the universal, $\omega^4$ regime deep inside the RSB phase, thereby shedding light on the origin of this universal behavior.

\section{The simplified KHGPS model}\label{simplified_khgps}

In the DIAG-like regime, the main effect of the random interactions $J_{ij}$ on the spectrum is to generate an effective distribution $P_m(m)$ of local stiffnesses $m$ (cf.~Eq.~(\ref{eq:eff_pot_KHGPS_m})), together with an effective random local field $H$ (cf.~Eq.~(\ref{eq:eff_pot_KHGPS_H})). We therefore consider a Hamiltonian describing a similar ensemble of anharmonic oscillators coupled through a mean-field ferromagnetic-like interaction:
\begin{equation}
    \mathcal{H}^S= \sum_{i} \left[\frac{A}{4!} x_i^4 + \frac{\kappa_i}{2} x_i^2 -(h+h_i) x_i \right]-\sum_{i,j} \frac{J}{2N} x_i x_j \quad .
\end{equation}
where the superscript ``S'' refers to the \textit{simplified} model. Here $h_i$ is a random external field, normally distributed around zero with variance $\sigma^2$, and $P(\kappa)$ is an arbitrary distribution of stiffnesses $\kappa$, which we again assume to have finite support bounded below by $\kappa_m$ and above by $\kappa_M$. Due to the low complexity of its energy landscape, this model cannot undergo an RS--RSB transition. Nevertheless, as we show below, its spectrum displays behavior similar to that observed in the KHGPS model: for $\kappa_m\!>\!0$, the spectrum resembles that found in the RS phase, whereas for $\kappa_m\!<\!0$ a crossover toward the universal $\omega^4$ spectrum emerges. Since replica-symmetry breaking is absent, the model can be studied analytically even in the presence of oscillators with negative stiffnesses and bistable local potentials.

\subsection{The replica trick}

We aim to determine the density of eigenvalues of the Hessian matrix of the simplified model, namely $\mathcal{M}_{ij}\!=\!-\frac{J}{N}\!+\!\delta_{ij}(\kappa_i\!+\!\frac{A}{2} x_i^2)$, in order to obtain the density of states, $\mathcal{D}(\omega)$. The matrix $\mathcal{M}$ can be decomposed into a constant matrix $\mathcal{C}$ with elements $\left(-\frac{J}{N}\right)$ and a diagonal matrix, which we denote $\mathcal{B}$ to better distinguish it from the diagonal matrix $\mathcal{A}$ of the original model. $\mathcal{B}$ features diagonal elements $b_i\!=\!\kappa_i\!+\!\frac{A}{2} x_i^2$. In the thermodynamic limit only the diagonal component contributes to the spectrum, so  the distribution of eigenvalues of the Hessian $\mathcal{M}$, $\rho_\mathcal{M}(\lambda)$, is fully determined by the spectrum of the diagonal matrix, $\rho_\mathcal{B}(b_i)$. Considering that $P(\kappa)$ has a finite support, one can show (see Appendix \ref{App:spectrum_hessian}) that
\begin{widetext}
\begin{equation}\label{prob_density_lambda}
    \rho_\mathcal{B}(b_i)=\int_{\kappa_{m}}^{\min(b_i,\kappa_{M})} d\kappa \, \frac{1}{\sqrt{2 A (b_i-\kappa)}} \left(\mathcal{P}_i\left(\sqrt{\frac{2}{A} (b_i-\kappa)},\kappa\right)+ \mathcal{P}_i\left(-\sqrt{\frac{2}{A}(b_i-\kappa)},\kappa\right)\right) \quad ,
\end{equation}
\end{widetext}
where ${\mathcal{P}_i}(\tilde{x},\tilde{\kappa})$ represents the joint probability of finding an oscillator with $\kappa_i\!=\!\tilde{\kappa}$ and $x_i\!=\!\tilde{x}$. We can immediately drop the index $i$ since it is evident from the Hamiltonian that the system is symmetric under permutation of indices, so ${\mathcal{P}_i}(\tilde{x},\tilde{\kappa})\!=\!{\mathcal{P}}(\tilde{x},\tilde{\kappa})$ for all $i$ and $\rho_{\mathcal{B}}(b_i)\!=\!\rho_{\mathcal{B}}(b)$. To compute ${\mathcal{P}}(\tilde{x},\tilde{\kappa})$ we can introduce a local auxiliary field $g(x_i, \kappa_i)$ for each soft spin, and compute the derivative of the logarithm of the partition function (see Appendix \ref{App:joint_prob}) 
\begin{equation}\label{distriP}
    \mathcal{P}(\tilde{x},\tilde{\kappa})=\left.-\frac{1}{\beta N}\frac{\delta }{\delta g(\tilde{x},\tilde{\kappa})} \langle \log(\mathcal{Z})\rangle\right|_{g=0}.
\end{equation}
where $\mathcal{Z}\!=\!\int \Pi_{i}dx_i   e^{-\beta \mathcal{H}^S[x]} $ and where $\langle \bullet \rangle$ represents averaging over the disorder.

We can make use of the replica trick to compute $\langle \log(\mathcal{Z})\rangle\!=\!\lim_{n \to 0} \frac{d \langle \mathcal{Z}^n \rangle}{d n}$ (see Appendix \ref{App:replica} for the details of the calculations for the replicated partition function), which yields
\begin{multline}\label{eq:logZ}
 \langle \log(\mathcal{Z})\rangle= - N \frac{\beta J}{2} z_0^2 \, + \\ N\int \!d\kappa\, ds \, \frac{e^{-s^2/2}}{\sqrt{2\pi}}P(\kappa)  \log\left( \int dx \, e^{ - \beta V_{\rm eff}(x, \kappa)}  \right)\,,
\end{multline}
with the effective local potential
\begin{equation}\label{eq:eff_pot}
    V_{\rm eff}(x,\kappa)=\frac{A}{4!} x^4  +  \frac{\kappa}{2}x^2 - ( J z_0 + h + s  \sigma ) x\,,
\end{equation}
where $s$ is a Gaussian random variable with zero mean and unit variance, and $z_0$ is a parameter that needs to be self-consistently determined through the following equation
\begin{equation}\label{eq:self_cons}
z_0= \int  d\kappa ds\, \frac{e^{-s^2/2}}{\sqrt{2\pi}}P(\kappa)   \frac{\int dx \, x e^{  - \beta V_{\rm eff}(x, \kappa)}  }{\int dx \, e^{ - \beta V_{\rm eff}(x, \kappa)} } \quad . 
\end{equation}
Finally, the joint probability ${\mathcal{P}}(\tilde{x},\tilde{\kappa})$ reads
\begin{equation}\label{eq:dist2}
    \mathcal{P}(\tilde{x},\tilde{\kappa})=\int  ds\, \frac{e^{-s^2/2}}{\sqrt{2\pi}}P(\tilde{\kappa}) \frac{  e^{  -\beta V_{\rm eff}(\tilde{x},\tilde{\kappa})}  }{ \int dx' \, e^{ -\beta V_{\rm eff}(x',\tilde{\kappa})} }
\end{equation}
Note that, since the couplings in the simplified system $\mathcal{H}^S$ are uniform rather than random, the calculation does not require any overlap matrix $\mathcal{Q}_{ab}$ between replicas (which do not interact with each other, see Appendix~\ref{App:replica}), and the replica trick is just a convenient way of performing the disorder average. Hence, there is no replica symmetry that can be broken, and the notions of RS and RSB phases do not apply to the simplified model.

\subsection{Zero-temperature limit}\label{zeroTlimit}

%%%%%%%%%%%%%%%%%%%%%%%%%%%
\begin{figure}
\centering
\vspace{0.1cm}{\includegraphics[scale=0.53]{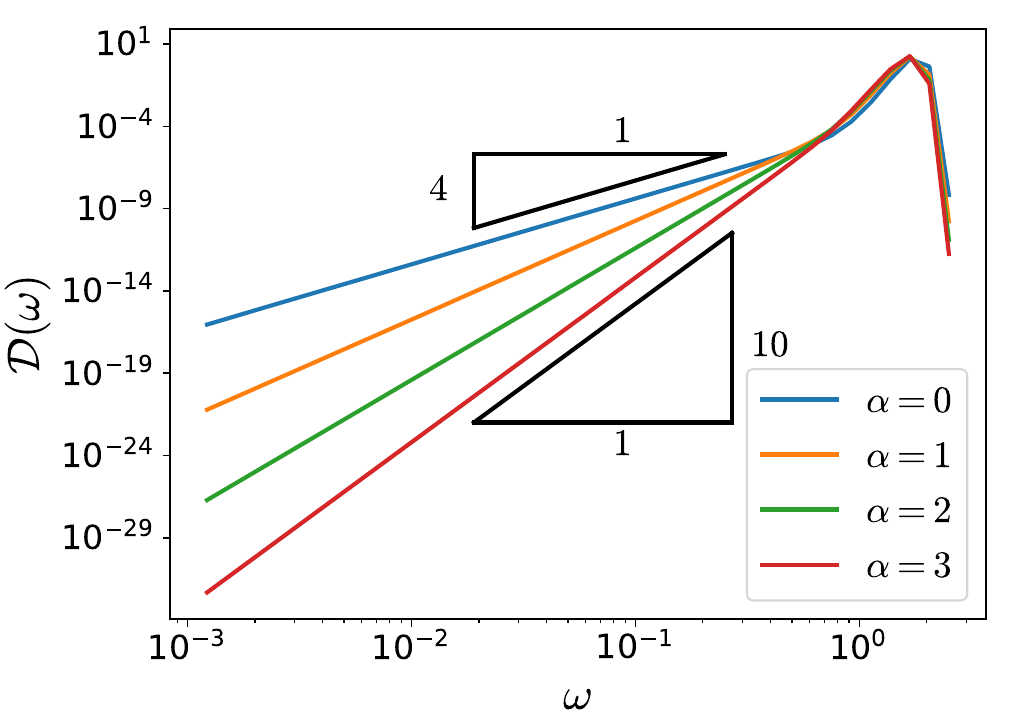}}
\caption{Density of states $\mathcal{D}(\omega)$ of the simplified KHGPS model for different values of the exponent $\alpha$, obtained by integrating numerically Eq.~\eqref{eq:dos_simple_KHGPS}. The lower bound of the stiffness distribution support is fixed to $\kappa_m\!=\!0$, while its upper bound is $\kappa_M\!=\!1$. All the other parameters of the system are also set to unity: $J\!=\!h\!=\!\sigma\!=\!A\!=\!1$. It is possible to appreciate the $\alpha$-dependent scaling regime of the density of states at low frequencies, which follows $\mathcal{D}(\omega)\!\sim\!\omega^{4 + 2\alpha}$.}
\label{fig:KHGPSvarDos}
\end{figure}
%%%%%%%%%%%%%%%%%%%%%%%%%%%

We consider now the low-temperature limit, $\beta\!\to\!\infty$, of the expressions \eqref{eq:self_cons} and \eqref{eq:dist2}. This allows us to compute the integrals over the variable $x$ through a saddle-point approximation. The self-consistency equation now reads
\begin{equation}\label{eq_zeroTemp_z}
    z_0= \int d\kappa \,P(\kappa)\int ds \, \frac{e^{-s^2/2}}{\sqrt{2\pi}}   x_0\,,
\end{equation}
where $x_0\!=\!x_0 (\kappa, s ; z_0)$ is the global minimum of the effective potential $\frac{d V_{\rm eff}(x,\kappa ; z_0)}{dx}|_{x=x_0}\!=\!0$. The equation for the distribution $\mathcal{P}(\tilde{x},\tilde{\kappa})$ is, instead, given by
\begin{equation}\label{eq:p_x_k}
    \mathcal{P}(\tilde{x},\tilde{\kappa})=P(\tilde{\kappa})\int ds \,   \frac{e^{-s^2/2}}{\sqrt{2\pi}} \delta(\tilde{x} - x_0)
\end{equation}
We can now make use of the composition property of the delta function
\begin{equation}\label{delta_property}
    \delta(f(s)) = \sum_i \frac{\delta(s- s_i)}{|f'(s_i)|}
\end{equation}
with $f(s)\!=\!\tilde{x}\!-\!x_0(s;\tilde{\kappa}, z_0)$ and where the prime denotes a derivative with respect to $s$ (see again Appendix \ref{App:replica} for details on the calculations). We obtain that the sum in Eq.~\eqref{delta_property} is composed of a single element $s_0$, namely
\begin{equation}
    s_0(x_0=\tilde{x})= \left(\frac{A}{3!} \tilde{x}^3 + \tilde{\kappa} \tilde{x} - (J z_0 + h ) \right)/\sigma\,,
\end{equation}
and the integral in Eq.~\eqref{eq:p_x_k} reduces to
\begin{equation}\label{eq:p_x_k_lowT}
    \mathcal{P}(\tilde{x},\tilde{\kappa})= P(\tilde{\kappa}) \frac{e^{-s_0^2/2}}{\sqrt{2\pi}|f'(s_0)|} \quad .
\end{equation}

We are interested in the low-energy spectrum of excitations, $\lambda\!\ll\!1$. In this limit, we obtain the expression for the density of eigenvalues by plugging expression \eqref{eq:p_x_k_lowT} into Eq.~\eqref{prob_density_lambda} and using Eq.~\eqref{eq:stiffness_distribution} for the distribution of stiffness $P(\kappa)$. One gets
\begin{widetext}
\begin{equation}\label{eq:dos_simple_KHGPS}
    \rho_\mathcal{M}(\lambda) \approx \frac{\lambda^{3/2}}{\sigma\sqrt{A \pi}}  \frac{(\alpha + 1)}{(\kappa_{M} - \kappa_{m})^{\alpha + 1}} \exp\left({-\frac{(J z_0 + h)^2}{2\sigma^2}}\right) \int_{\max(- \frac{1}{2}, \kappa_{m}/\lambda)}^{\min(1,\kappa_{M}/\lambda)}  d\kappa \, \frac{(\lambda\kappa - \kappa_{m})^\alpha}{\sqrt{1 - \kappa}} \quad ,
\end{equation}    
\end{widetext}
where we rescaled the variable $\kappa\!\to\!\lambda\kappa$ (see Appendix \ref{app:Dos_var} for details of the calculation).

Notice that Eq.~\eqref{eq:dos_simple_KHGPS} is structurally analogous to the curvature distribution of the original KHGPS (cf.~the cumulative distribution $G (\tilde{a})$ given by Eq.~\eqref{eq:Gscaling}). Both of them, in fact, display an explicit exponential dependence on the control parameters $J$ and $h$, and a scaling with respect to the main variable given by the product of a fixed, constant power with exponent $\frac{3}{2}$ (note that it was $\frac{5}{2}$ for Eq.~\eqref{eq:Gscaling}, since there we are considering the cumulative) and an additional $\alpha$-dependent factor coming from the binomial term inside the integral. In the current case, the role of the effective stiffness $m$ in the original KHGPS model is played by $\kappa$, whose distribution support we can directly control.

We focus on the support of the spectrum and on its scaling with respect to $\lambda$. We observe again that when the stiffness distribution is gapped, $\kappa_m\!>\!0$, the eigenvalue distribution is also gapped. When $\kappa_m\!=\!0$, instead, the gap closes and we obtain that $\rho_\mathcal{M}\!\sim\!\lambda^{3/2 + \alpha}$, which yields $\mathcal{D}(\omega)\!\sim\!\omega^{4+2\alpha}$. The density of states is, thus, gapless and non-universal: its scaling is sensitive to the scaling of $P(\kappa)$ itself near its lower edge, in the same fashion as the original KHGPS model (see Fig.~\ref{fig:KHGPSvarDos}). Lastly, when $\kappa_m\!<\!0$ the binomial term $(\lambda \kappa\!-\!\kappa_m)^\alpha$ will give rise to multiple terms which contribute additively to $\rho_\mathcal{M}$. We devote the next section to studying in more depth this last case.

%%%%%%%%%%%%%%%%%%%%%%%%%%%
\begin{figure*}
\centering
{\includegraphics[scale=0.48]{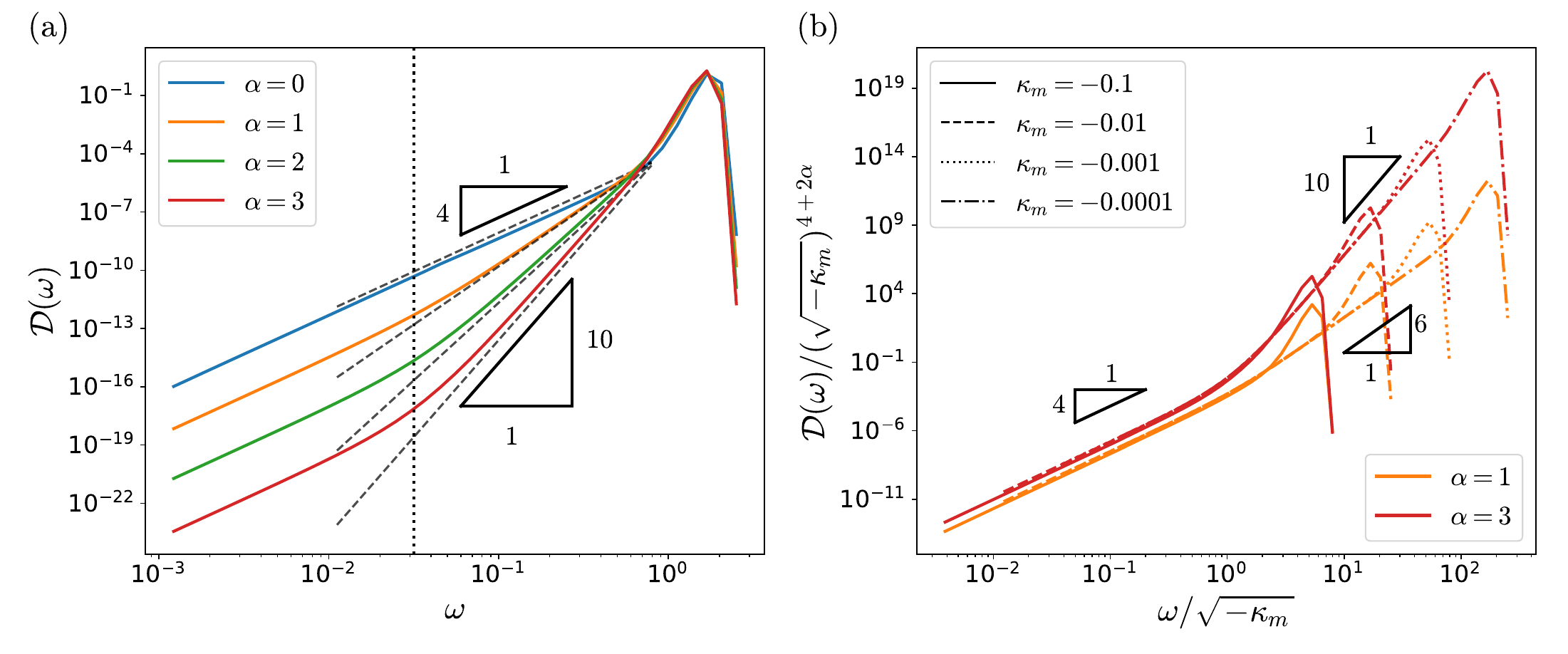}}
\caption{(a) Density of states of the simplified KHGPS model for different values of the exponent $\alpha$. The lower bound of the stiffness distribution is fixed to $\kappa_m\!=\!-0.001$, while the rest of the parameters are set to unity, as in the previous plot of Fig.~\ref{fig:KHGPSvarDos}. The two different scaling regimes, visible only for $\kappa_m\!<\!0$, are evident: at low frequencies the density of states scales like $\omega^4$, while at larger frequencies the $\alpha$-dependent scaling takes over (the gray dashed lines indicate the $\omega^{4 + 2\alpha}$ scaling). The vertical dotted line represents the crossover frequency, $\omega_{\text{cross}}\!\sim\!\sqrt{-\kappa_m}\!=\!\sqrt{0.001}$. (b) Rescaled density of states with respect to the crossover frequency for selected values of $\alpha$ (denoted by different colors) and of $\kappa_m$ (different line styles). When the stiffness distribution penetrates more into the negative axis, the $\alpha$-dependent scaling disappears in favor of the quartic scaling.}
\label{fig:KHGPSvarCross}
\end{figure*}
%%%%%%%%%%%%%%%%%%%%%%%%%%%

\section{Recovering the universal quartic scaling}\label{crossover}

\subsection{Crossover frequency in the simplified KHGPS}

We now inspect the regime in which the stiffness distribution extends into the negative axis, $\kappa_m\!<\!0$. To get the asymptotic scaling of $\rho_\mathcal{M}(\lambda)$ (cf.~Eq.~\eqref{eq:dos_simple_KHGPS}) at small $\lambda$, it is insightful to simply consider the binomial term inside the integral. It is, in fact, immediate to observe that the leading contribution in the binomial expansion of $(\lambda\kappa-\kappa_m)^\alpha$ will depend on the relative strength of $\lambda$ and $\kappa_m$. In particular, we have
\begin{equation}\label{eq:binomial_scaling}
    \left(\lambda \kappa -\kappa_m \right)^{\alpha} \sim \begin{cases}
        (-\kappa_m)^\alpha\sim \mathcal{O}(\lambda^0) \, &\text{if } \lambda \ll |\kappa_m| \\
        (\lambda \kappa)^\alpha \sim \mathcal{O}(\lambda^\alpha) \, &\text{if }  |\kappa_m| \ll \lambda \ll \kappa_M
    \end{cases} \, .
\end{equation}

Let us consider first the case $\lambda\!\ll\!|\kappa_m|$. Restricting ourselves to the leading order contribution, the integral in Eq.~\eqref{eq:dos_simple_KHGPS} can be approximated by
\begin{equation}
    ( - \kappa_{m})^\alpha\int_{- \frac{1}{2}}^{1}  d\kappa \, \frac{1}{\sqrt{1 - \kappa}} = \sqrt{6} (- \kappa_{m})^\alpha \, ,
\end{equation}
which leads to the following scaling of the spectrum
\begin{equation}\label{eq:scaling_1}
    \rho_\mathcal{M} \sim \lambda^{3/2} P_0 \, ,
\end{equation}
with $P_0\!\equiv\!P(\kappa\!=\!0)\!\sim\!(-\kappa_m)^\alpha$. This means that, near the origin, the spectrum is proportional to the density of oscillators at zero stiffness rather than to the scaling of $P(\kappa)$ near its lower edge, as it was when $\kappa_m\!=\!0$ (see Sec.~\ref{zeroTlimit}).

When $\lambda$ grows and becomes larger in magnitude than the lower bound, $\lambda\!\gg\!|\kappa_m|$, according to Eq.~\eqref{eq:binomial_scaling} the main contribution to the integral will come from the term of degree $\alpha$:
\begin{equation}
    \lambda^\alpha\int_{0}^{1}  d\kappa \, \frac{\kappa^\alpha}{\sqrt{1 - \kappa}} =  \lambda^\alpha B\left(\alpha+1,\frac{1}{2}\right) \, ,
\end{equation}
where $B(\alpha\!+\!1, \frac{1}{2})$ is the Beta function. Consequently, the spectrum will follow
\begin{equation}\label{eq:scaling_2}
    \rho_\mathcal{M}(\lambda) \sim \lambda^{3/2 + \alpha} \, ,  
\end{equation}
displaying again an $\alpha$-dependent scaling.

In terms of the density of states, this result translates to the following scaling
\begin{equation}
    \mathcal{D}(\omega) \sim \begin{cases}
        \omega^4 P_0 \quad &\text{for $\omega\!<\!\omega_{\text{cross}}$} \\
        \omega^{4 + 2\alpha} \quad &\text{for $\omega_{\text{cross}}\!<\!\omega\!<\!\sqrt{\kappa_{M}}$}
    \end{cases}
\end{equation}
with the crossover frequency obtained by equating the density of states in the two regimes, Eq.~\eqref{eq:scaling_1} and Eq.~\eqref{eq:scaling_2}. This gives
\begin{equation}\label{eq:crossover_freq_simple}
    \omega_{\text{cross}} \sim P_0^{\frac{1}{2\alpha}} \sim \sqrt{-\kappa_m}  \quad .
\end{equation}
This analytical result is confirmed by solving numerically Eq.~\eqref{eq:dos_simple_KHGPS}: we plot the results for different values of the exponent $\alpha$ and of the lower bound $\kappa_m$ in Fig.~\ref{fig:KHGPSvarCross}. A more detailed derivation (see Appendix \ref{app:Dos_var}) gives an $\alpha$-dependent crossover frequency, in particular
\begin{equation}
    \omega_{\text{cross}}=\sqrt{-\kappa_m} \left( \frac{\sqrt{6}}{B\left(\alpha + 1, \frac{1}{2}\right)}\right)^{\frac{1}{2\alpha}} \, ,
\end{equation}
showing that in the limit $\alpha\!\to\!0$ the quartic scaling takes over and the crossover frequency diverges (note that $B(1, \frac{1}{2})\!=\!2$ and so the expression in the parenthesis is larger than unity).

We want to stress a couple of important points, which are relevant to understand how to apply the results obtained from this simplified model to the original KHGPS system. (i) The first one is that when $\kappa_m\!=\!0$ we have $P_0\!=\!0$ for all the values of the exponent $\alpha$ except zero. In other words, for a uniform stiffness distribution the probability density at zero stiffness, $P_0$, is finite for any $\kappa_m\!\le\!0$, including $\kappa_m\!=\!0$, and therefore the spectrum is always quartic. (ii) The second one, instead, is that the span of the quartic regime is controlled by the extension of the support of $P(\kappa)$ into the negative axis: the more the distribution shifts to negative values, the larger the frequency at which the quartic tail crosses over to the $\alpha$-dependent scaling.

%%%%%%%%%%%%%%%%%%%%%%%%%%
\begin{figure*}
\centering
{\includegraphics[scale=0.5]{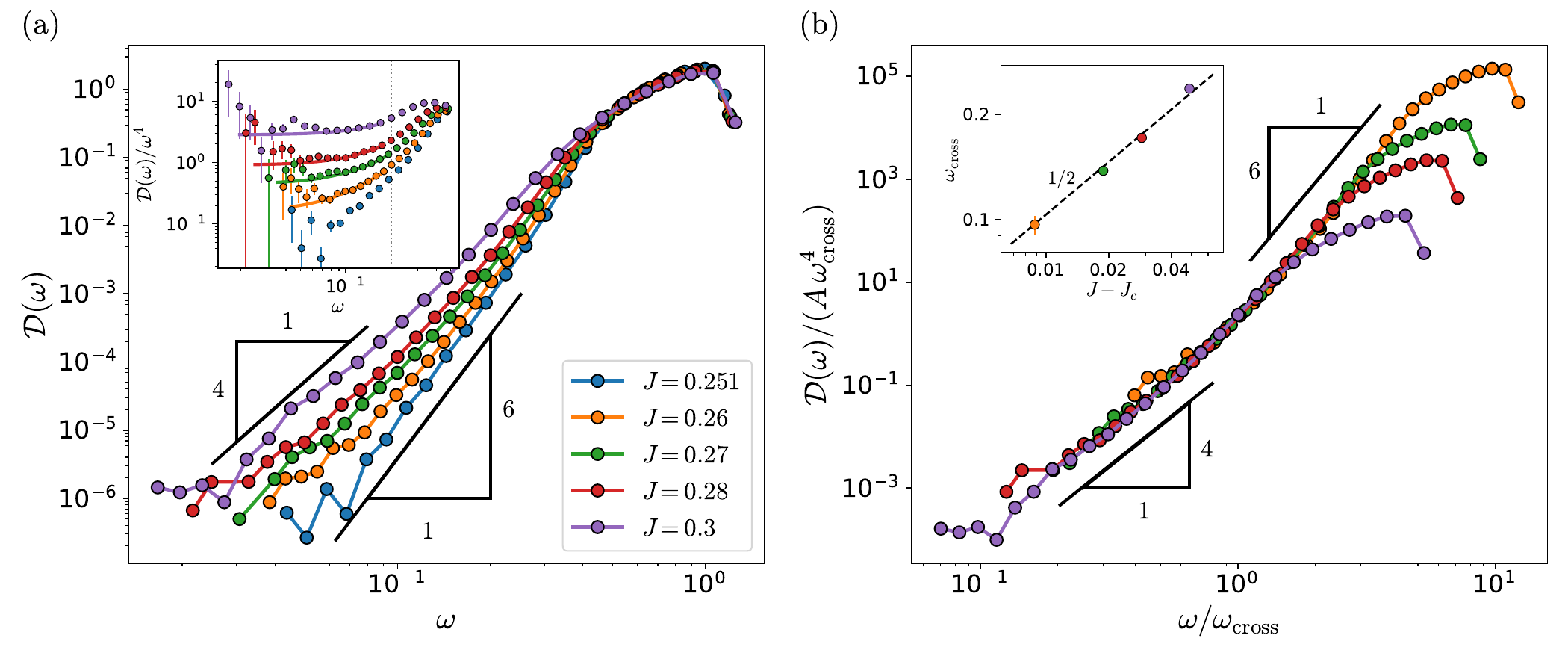}}
\caption{(a) Density of states of the KHGPS model for different values of $J$, computed for a simulated system of $N\!=\!10\,000$ oscillators with $\alpha\!=\!1$, $\kappa_m\!=\!0.1$ and $\kappa_M\!=\!1$. The data is averaged over $50\,000$ realizations. We fix the external field to $h\!=\!0.1975$, for which the critical coupling at which the system transitions to the RSB phase is $J_c\!=\!0.2512$ (the blue dataset in this plot corresponds to the one shown in Fig.~\ref{fig:phase_diagram} marked by the upper triangle $\triangle$; for the other datasets, we move horizontally in the $(J,h)$ phase diagram, entering inside the RSB phase while keeping $h$ fixed). The density of states on the critical line (blue dataset) shows the predicted scaling with $\omega^{4 +2\alpha}\!=\!\omega^6$. By entering deeper and deeper in the RSB phase we observe the predicted growth of a low-frequency quartic scaling which gradually takes over (the purple dataset at $J\!=\!0.3$ shows almost no sign of the $\omega^{6}$ scaling). The inset shows the same densities of states, rescaled by $\omega^4$ and fitted with the function $f(\omega)\!=\!A\omega^4\!+\!B\omega^6$. (b) Rescaled density of states with respect to $\omega_{\text{cross}}\!=\!(A/B)^{1/(2\alpha)}\!=\!\sqrt{A/B}$, whose scaling with respect to the distance from the transition line is shown in the inset (the color code is the same as in panel (a)) and confirms that $\omega_{\text{cross}}\!\sim\!\sqrt{J-J_c}$.}
\label{fig:crossoverKHGPS}
\end{figure*}
%%%%%%%%%%%%%%%%%%%%%%%%%%

\subsection{Crossover frequency in the original KHGPS}

Finally, we return to the original KHGPS model. We now address the last point left open in Sec.~\ref{KHGPS_non_universal}, that is the case when the lower bound of the effective stiffness distribution becomes negative, $m_m\!<\!0$. We already mentioned the fact that, even if the replica-symmetric ansatz breaks down, and hence Eq.~\eqref{eq:Gscaling} is no longer valid, a description of the system in terms of a collection of independent oscillators in the effective potential of Eq.~\eqref{eq:eff_pot_KHGPS} with effective fields $H$ (given by a different, $m$-dependent, distribution) and renormalized stiffness $m\!=\!\kappa\!-\!J^2\chi$ (where $\chi$ is now the susceptibility of the fullRSB solution) is still possible \cite{folena_marginal_2022}. For this reason we expect that when the lower bound of the effective stiffness distribution crosses over to negative values, the same mechanism that we just spelled out for the simplified KHGPS will also take place in the original KHGPS model. In particular, we will have multiple terms contributing to the curvature distribution, with the low-curvature limit governed by the density of oscillators at zero effective stiffness, $P_m(0)\!\propto\!(-m_m)^\alpha$, yielding $P(\tilde{a})\!\sim\!\tilde{a}^{3/2}(-m_m)^\alpha$. This eventually gives rise to the universal and $\alpha$-independent density of states $\mathcal{D}(\omega)\!\sim\!\omega^4$. At larger curvature (and, consequently, at larger frequency), the other terms will take control of the scaling of the spectrum, producing a crossover to $\mathcal{D}(\omega)\!\sim\!\omega^{4 + 2\alpha}$. As for the simplified model, we expect to find a crossover frequency $\omega_{\text{cross}}$ between the two regimes, which sets the scale below which the universal quartic distribution of QLM frequencies is observed. We suppose that, also in this case, $\omega_{\text{cross}}$ will depend on how deep the stiffness distribution has moved into the negative axis, hence $\omega_{\text{cross}} \sim \sqrt{-m_m}$.

The lower bound $m_m$ of the shifted stiffness distribution is not an input parameter, but rather an emergent property. Its scaling with respect to $J$ and $h$, according to its definition as given in Eq.~\eqref{eq:eff_pot_KHGPS_m}, depends on the quantity $J^2\chi$, i.e.\ on the disorder-induced shift of the stiffness distribution, already identified in Refs.~\cite{rainone_mean-field_2021, folena_marginal_2022, suda_local_2026} as a relevant stiffness scale: oscillators with $\kappa\!<\!J^2\chi$, i.e.\ with $m\!<\!0$, have double-well effective potentials~\cite{folena_marginal_2022} and are stabilized only by their quench-induced displacement~\cite{suda_local_2026}. These oscillators appear only inside the DIAG-like RSB phase, and the width of their stiffness range, $-m_m\!=\!J^2\chi\!-\!\kappa_m$, controls the extension of the quartic scaling regime. Since $\chi$, and hence $m_m$, is not known explicitly in the RSB phase, we assume continuity of the susceptibility $\chi$ and of its derivative across the RSB transition in order to obtain the scaling of $m_m$ with respect to $J$. We then test this prediction numerically. For a fixed field $h$, a Taylor expansion of $m_m$ around a point on the transition line yields a linear dependence on the distance, $-m_m\!\sim\!J\!-\!J_c(h)$, given that $\chi$ and its derivative with respect to $J$ stay finite at the DIAG-like RSB transition. For the crossover frequency, this implies
\begin{equation}
    \omega_{\text{cross}} \sim \sqrt{J-J_c}
\end{equation}
We confirm the presence of a crossover in the density of states and the scaling of the crossover frequency $\omega_{\text{cross}}$ with respect to $J$ in Fig.~\ref{fig:crossoverKHGPS}, where we plot the spectrum of an ensemble of simulated KHGPS systems of $N\!=\!10\,000$ oscillators computed at an external field $h\!=\!0.1975$ and with different values of the coupling $J$, starting from the transition line (where the coupling assumes the critical value $J_c\!=\!0.2512$) and ending deep inside the RSB phase. The samples are quenched from an initial non-zero random configuration of the oscillators, with each $x_i$ uniform in $[-\varepsilon/2, \varepsilon/2]$ and $\varepsilon\!=\!10^{-2}$ as in Refs.~\cite{suda_yielding_2025, suda_local_2026}, using a conjugate gradient algorithm. The function $f(\omega)\!=\!A\omega^4\!+\!B\omega^{4+2\alpha}$, with $A$ and $B$ parameter-dependent coefficients, is fitted to the densities of states obtained for different values of $J$. The crossover frequency $\omega_{\text{cross}}\!\sim\!\sqrt{-m_m}$ is extracted as $\omega_{\text{cross}}\!=\!(A/B)^{\frac{1}{2\alpha}}$, and its scaling is confirmed to follow $\omega_{\text{cross}}\!\sim\!\sqrt{J-J_c}$.

To summarize, we found that the original KHGPS model in the RSB phase shows the same behavior observed in its simplified version: a density of states that depends on the density of oscillators at zero effective stiffness, $P_m(0)$ (equivalent to $P_0$ in the simplified KHGPS), but not on the fine details of the (input) stiffness distribution $P(\kappa)$. Whenever the disorder-induced shift $(-J^2\chi)$ in the original stiffness distribution (initially with a positive support) increases $P_m(0)$ to a finite value, the density of states develops a quartic scaling $\mathcal{D}(\omega)\!\simeq\!A\,\omega^4$ for $\omega\!\lesssim\!\omega_{\rm cross}$. Moreover, above the crossover frequency $\omega_{\rm cross}\!\sim\!\sqrt{-m_m}$, set by the same shift, the quartic law crosses over to the non-universal scaling $\mathcal{D}(\omega)\!\sim\!\omega^{4+2\alpha}$ found on the transition line, Eq.~\eqref{eq:alpha-dependent-vdos}, in which the exponent $\alpha$ of the lower edge of the stiffness distribution, $P_m(m)\!\sim\!(m\!-\!m_m)^\alpha$, reappears. Since $\omega_{\rm cross}\!\sim\!(J\!-\!J_c)^{1/2}$ grows with the distance from the transition line, the quartic regime covers an increasingly large portion of the low-frequency spectrum as one moves deeper into the RSB phase.

\section{Discussion}\label{discussion}

%Note that by imposing less simple shapes of the stiffness distribution it would be possible to have re-entrances in the RSB phase. E.g. if $P(\kappa)$ is bimodal and goes to zero in the middle, then when the stiffness distribution is shifted we can reach a point in which $P(m=0)=0$, obtaining again an $\alpha$-dependent density of states.

In the present work we have studied how the low-frequency vibrational spectrum of the KHGPS model depends on details of the input distribution $P(\kappa)$ of local stiffnesses, focusing in particular on the DIAG-like regime, near and beyond the RS--RSB transition. In this regime the lower edge of the spectrum of the Hessian is controlled by its diagonal part, $\rho_\mathcal{M}(\lambda)\!\sim\!(\lambda\!-\!\lambda_m)^\nu$ (the spectrum becomes gapless, $\lambda_m\!=\!0$, at the transition line), with the same exponent $\nu$ as the distribution of the diagonal elements, $P(a)\!\sim\!(a\!-\!a_m)^\nu$. The diagonal elements, $a_i\!=\!\kappa_i\!+\!\frac{1}{2}(x_i^*)^2$, contain the input stiffnesses explicitly, and $P(a)$ near its lower edge inherits the behavior of $P(\kappa)$ near $\kappa_m$ through the effective stiffness distribution $P_m(m)\!=\!P(m\!+\!J^2\chi)$, cf.~Eq.~\eqref{eq:Gscaling}. Therefore the low-frequency spectrum in the DIAG-like regime becomes dependent on the details of $P(\kappa)$, whereas in the GOE-like regime its edge is the universal semicircular one, $\rho_\mathcal{M}\!\sim\!(\lambda\!-\!\lambda_m)^{1/2}$, independent of $P(\kappa)$.

We showed how the input stiffness distribution shapes the phase diagram, choosing the power-law family $P(\kappa)\!\sim\!(\kappa\!-\!\kappa_m)^\alpha$ as our working example. We observed that increasing $\alpha$ shifts the RSB phase boundary to larger disorder $J$ and drives the special point, which separates the continuous (spin-glass-like, with diverging susceptibility and $\mathcal{D}(\omega)\!\sim\!\omega^2$) from the discontinuous branch of the transition line, towards smaller values of the external field $h$. We obtained in closed form the exponent $\alpha_0\!=\!w/(w\!-\!\kappa_m)$ at which the special point reaches $h\!=\!0$, with $w$ the width of the support of $P(\kappa)$. For $\alpha\!>\!\alpha_0$ (and a sufficiently large width of the stiffness distribution) the continuous transition disappears altogether: replica symmetry can only be broken discontinuously, through the appearance of bistable effective potentials, even in zero field. This differs from the gapless case $\kappa_m\!=\!0$ considered in Ref.~\cite{rainone_mean-field_2021}, where at $h\!=\!0$ there is no RS phase at all and any coupling $J\!>\!0$ shifts the lower edge of $P_m$ to $m_m\!=\!-J^2\chi\!<\!0$, so that $J_c(0)\!=\!0$ and the system is always in its DIAG-like regime for any non-zero coupling $J$. For $\kappa_m\!>\!0$, instead, the RS phase survives at $h\!=\!0$ up to a finite coupling, and for $\alpha\!>\!\alpha_0$ this coupling becomes $J_c(0)\!=\!\sqrt{\kappa_m w\,\alpha/(\alpha\!+\!1)}$, at which $m_m$ reaches zero while $\lambda_R$ is still positive, and the whole transition line $J_c(h)$, down to $h\!=\!0$, belongs to the discontinuous branch. In this case the continuous, spin-glass-like transition does not occur at any value of the field, showing that whether the KHGPS model displays a spin-glass-like transition is not a generic property of the model, but depends on the statistics of the local stiffnesses.

Moreover, in our work we showed analytically that for a power-law--distributed stiffness, at the discontinuous RS--RSB transition, where the lower bound $m_m$ of the support of the effective stiffness distribution $P_m(m)$ just reaches zero, the curvature distribution becomes $\alpha$-dependent and (for $h\!\ne\!0$) scales as $P(\tilde{a})\!\sim\!\tilde{a}^{3/2+\alpha}$, and hence $\mathcal{D}(\omega)\!\sim\!\omega^{4+2\alpha}$, which we confirmed by direct numerical diagonalization of the Hessian of large KHGPS realizations. The quartic spectrum previously reported at the boundary of the RS phase is therefore the special case $\alpha\!=\!0$ of a non-universal family, and not a generic property of the transition.

Universality is, instead, restored once we move inside the RSB phase. As the coupling $J$ increases beyond $J_c(h)$, the shift $(-J^2\chi)$ pushes the lower edge of $P_m$ to negative values and a finite density of oscillators at vanishing effective stiffness appears, $P_m(0)\!\propto\!(-m_m)^\alpha$. Due to the presence of these zero-stiffness oscillators, a quartic regime $\mathcal{D}(\omega)\!\sim\!P_m(0)\,\omega^4$ develops at low frequency, independently of $\alpha$, and extends up to a certain frequency scale, $\omega_{\text{cross}}$, before crossing back over to the $\alpha$-dependent scaling. The simplified KHGPS model, which lacks replica-symmetry breaking but reproduces the effective \textit{uncoupled} oscillator physics, allowed us to derive this crossover explicitly and to identify the extension of the stiffness distribution into the negative axis as its relevant scale, $\omega_{\text{cross}}\!\sim\!\sqrt{-\kappa_m}$: below $\omega_{\text{cross}}$ the spectrum is quartic, above it the $\alpha$-dependent scaling $\sim\!\omega^{4+2\alpha}$ survives. In the KHGPS model the same scale is set by $\sqrt{-m_m}$, which we have shown numerically to follow $\omega_{\text{cross}}\!\sim\!\left(J\!-\!J_c(h)\right)^{1/2}$. The crossover scale grows with the distance from the transition line, so that the quartic regime soon covers the entire low-frequency tail. This supports the picture that structural glasses, which robustly and universally display $\mathcal{D}(\omega)\!\sim\!\omega^4$~\cite{lerner_low-energy_2021}, reside deep inside a marginal phase characterized by full replica-symmetry breaking, rather than close to its boundary.

In conclusion, our results showed that the low-frequency scaling of $\mathcal{D}(\omega)$ is governed by the weight that the \textit{effective} stiffness distribution $P_m(m)$, i.e.\ the bare distribution $P(\kappa)$ shifted by the disorder-induced term $(-J^2\chi)$, retains at zero stiffness. When $P_m(0)$ is finite, the quartic law emerges, with an amplitude proportional to $P_m(0)$; when it vanishes, the spectrum inherits the shape of $P(\kappa)$ near its lower edge. Interpreting $P(\kappa)$ as the spectrum of local excitations of the parent liquid before the quench, our results suggest that disordered systems close to the RS--RSB transition can preserve in their spectrum information about the original microscopic details, which are instead erased deep inside the glassy phase, where only the amplitude of the quartic tail, through $P_m(0)$, retains memory of them.

%Several questions remain open. A quantitative theory of the crossover in the KHGPS model requires the solution of the full RSB equations, and in particular the behavior of the effective stiffness and field distributions in the marginal phase~\cite{folena_marginal_2022}. It would also be interesting to relate the dependence of the amplitude of the quartic tail on $P_m(0)$ to its known dependence on glass preparation in simulations~\cite{wang_low-frequency_2019, rainone_pinching_2020}, and to explore the consequences of our findings for two-level systems in the quantized version of the model~\cite{maimbourg_two-level_2024}.

\acknowledgments
\vspace{-0.2cm}
This work was supported by the Institute of Physics of the University of Amsterdam. G.D. acknowledges funding from ANID FONDECYT No 1251982. We thank Eran Bouchbinder and Makoto Suda for their comments on the manuscript.

\begin{widetext}

\appendix

\section{The KHGPS model}

\subsection{The spectrum}\label{app:spectrum_KHGPS}

To find the spectrum of the Hessian matrix $\mathcal{M}$ we consider the \textit{resolvent} and its $N\!\to\!\infty$ limit
\begin{equation}
    \mathfrak{g}_{\mathcal{M}} (z) = \lim_{N \to \infty} \frac{1}{N} \Tr( z \mathbf{1} - \mathcal{M})^{-1} = \int d \lambda \, \frac{\rho_{\mathcal{M}}(\lambda)}{z - \lambda} \quad .
\end{equation}
From the resolvent, the spectrum of the Hessian is readily obtained as
\begin{equation}\label{eq:spectrum}
    \rho_{\mathcal{M}} (\lambda) = \frac{1}{\pi} \lim_{\eta \to 0^+} \Im \left( \mathfrak{g}_{\mathcal{M}} (\lambda - i \eta) \right) \quad .
\end{equation}
Since $\mathcal{M}$ is the sum of a diagonal matrix $\mathcal{A}$, with elements $a_i\!=\!\kappa_i\!+\!\frac{1}{2}(x_i^*)^2$, and of a matrix in the GOE ensemble, its limiting spectrum satisfies the self-consistent Pastur equation
\cite{pastur_eigenvalue_2011, voiculescu_addition_1986} (see also \cite{bun_cleaning_2017}, § 2.3.2)
\begin{equation}
    \mathfrak{g}_{\mathcal{M}}(z)\;=\;\mathfrak{g}_{\mathcal{A}}\!\big(z-J^2 \mathfrak{g}_{\mathcal{M}}(z)\big).
\label{eq:selfcons}
\end{equation}
where $J^2$ is the variance of the elements of the Gaussian matrix and $\mathfrak{g}_{\mathcal{A}} (z)$ is the resolvent of the diagonal matrix. The Pastur equation yields
\begin{equation}\label{eq:resolvent_again}
    \mathfrak{g}_{\mathcal{M}} (z) = \int_{a_m}^{a_M} da \, P(a) \frac{1}{z - J^2\mathfrak{g}_{\mathcal{M}} (z) - a}
\end{equation}
We can now perform the following substitution
\begin{equation}\label{eq:substitution}
    \mathfrak{g}_{\mathcal{M}} (z) = g(z) + \frac{z}{J^2}
\end{equation}
and isolate $z$ in Eq.~\eqref{eq:resolvent_again}: 
\begin{equation}\label{eq:f_function}
    z = - J^2 \int_{a_m}^{a_M} da \, P(a) \left( g + \frac{1}{a + J^2g} \right) \equiv \mathcal{F}(g) \quad .
\end{equation}
The entire problem of finding the spectrum and its edges is thereby reduced to analyzing \eqref{eq:f_function}: for real $z$ outside the spectrum, $z\!=\!\mathcal{F}(g)$ admits a real branch $g(z)$ and $\mathfrak{g}_{\mathcal{M}}(z)\!=\!\mathfrak{g}_{\mathcal{A}}(g(z))$ remains real-analytic. The subset of $\mathbb{R}$ that is complementary to the image of the real-analytic branch of $\mathcal{F}$ constitutes the support of the spectrum of the Hessian.

To find the spectral edges, then, we start by assuming that the probability distribution of the diagonal elements $P(a)$ has a support limited to the interval $[a_m, a_M]$. Then $\mathcal{F}(g)$ exists and is real when
\begin{equation}
    a + J^2 g\ne 0 \quad \forall \, a \in [a_m, a_M] \quad .
\end{equation}
Therefore, the domain of $\mathcal{F}$ is $\mathcal{D}\!=\!\mathbb{R} \setminus \left[-\frac{a_M}{J^2}, -\frac{a_m}{J^2} \right]$, and the support of the spectrum is given by
\begin{equation}
    \Supp (\rho_{\mathcal{M}}) = \mathbb{R} \setminus \mathcal{F}(\mathcal{D}) \equiv [\lambda_m, \lambda_M]
\end{equation}
(we assumed for simplicity that the support is a connected set, the interval limited by $\lambda_m$ and $\lambda_M$, but we will later confirm that this is indeed the case). To obtain $\lambda_m$ and $\lambda_M$ we need to study $\mathcal{F}$ and to find its minimum and maximum. The first derivative of $\mathcal{F}(g)$ is
\begin{equation}\label{eq:f_first_derivative}
    \mathcal{F}'(g) = - J^2 \int_{a_m}^{a_M} da \, P(a) \left( 1 - \frac{J^2}{(a + J^2g)^2} \right)
\end{equation}
while the second derivative reads
\begin{equation}\label{eq:second_derivativeF}
    \mathcal{F}''(g) = - 2 J^6 \int_{a_m}^{a_M} da \, P(a) \frac{1}{(a + J^2g)^3} 
\end{equation}
From the sign of Eq.~\eqref{eq:second_derivativeF} we know that $\mathcal{F}$ is always convex for $g\!\leq\!- \frac{a_M}{J^2}$ (hence features at most a single minimum $g_M$, such that $\mathcal{F}(g_M)\!=\!\lambda_M$, the upper edge of the support) and similarly that $\mathcal{F}$ is always concave for $g\!\geq\!- \frac{a_m}{J^2}$ (with at most a single maximum $g_m$, such that $\mathcal{F}(g_m)\!=\!\lambda_m$, the lower edge of the support). See Fig.~\ref{fig:F(g)} for a qualitative sketch of the function. Note that this is self-consistent with our initial assumption of a connected support.

We will now focus our attention on the maximum $g_m$, which gives us the lower edge $\lambda_m$ of the support of the spectrum. The study of the other edge is identical. First of all, we notice that the maximum may not exist, in the sense that there might not be any value of $g\!\geq\!- \frac{a_m}{J^2}$ such that $\mathcal{F}'(g)\!=\!0$. In this case, the $g_m$ that determines the lower edge $\lambda_m$ is exactly $g_m\!=\!- \frac{a_m}{J^2}$. The limiting case between the existence and the absence of the maximum is given by the condition $\mathcal{F}'\left(g_m\!=\!-\frac{a_m}{J^2}\right)\!=\!0$, which corresponds to the maximum falling exactly on the edge of the domain. We can, then, easily distinguish the two cases by checking the sign of the first derivative on the edge, $\mathcal{F}'\left(-\frac{a_m}{J^2}\right)$. Let us define the quantity
\begin{equation}\label{app:eq:Lambda_def}
    \Lambda \equiv -\frac{1}{J^2} \mathcal{F}'\left(-\frac{a_m}{J^2}\right) = 1 - J^2\int_{a_m}^{a_M} da \, P(a) \frac{1}{(a - a_m)^2} \quad .
\end{equation}
If $\Lambda\!<\!0$ the maximum exists, while if $\Lambda\!>\!0$ the maximum does not exist (see Fig.~\ref{fig:F(g)} for a qualitative graphic description). The condition $\Lambda\!=\!0$ separates the two cases.

%%%%%%%%%%%%%%%%%%%%%%%%%%%%
\begin{figure*}
\centering
{\includegraphics[scale=0.55]{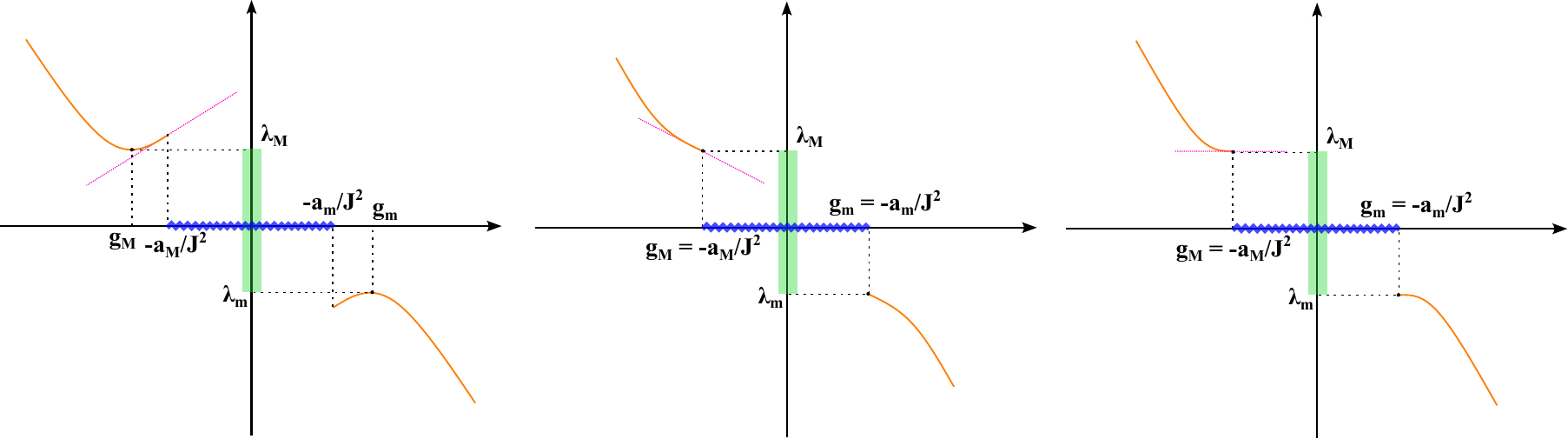}}
\caption{Qualitative sketch of the function $\mathcal{F}(g)$ (in orange) for different values of the coupling $J$. The highlighted blue interval on the x-axis is where the function is not defined for real values, while the green interval on the y-axis is the forbidden region for $g$ real and represents the support of the spectrum, $\Supp(\rho_{\mathcal{M}})$. Note that for different values of $J$ the function shows a maximum (minimum) where its first derivative is zero (leftmost panel), or it does not (central panel). A critical value of the coupling, for which the maximum (minimum) falls exactly on the edge of the domain, $g_m\!=\!- \frac{a_m}{J^2}$, separates the two cases (rightmost panel). The sign of $\mathcal{F}'(g)$ evaluated on the edge of the domain (thin pink dotted line in the plot, drawn only for the edge near the minimum for simplicity) allows us to distinguish the two cases.}
\label{fig:F(g)}
\end{figure*}
%%%%%%%%%%%%%%%%%%%%%%%%%%%%%%

Now that we know the limits of the support of the spectrum, $\lambda_m$ and $\lambda_M$, and for which value of $g$ they occur, we can expand the function $\mathcal{F}(g)$ around them. Let us consider each case separately, starting from the one where the maximum (minimum) exists.

\subsubsection*{Case 1: GOE-like spectrum}

Let us expand the function around the lower edge of the support of the spectrum, $(g_m, \lambda_m\!=\!\mathcal{F}(g_m))$. We take e.g. $z\!=\!\lambda_m\!+\!\delta \lambda$ and $g\!=\!g_m\!+\!\delta g$, with $\delta \lambda$ and $\delta g$ both complex. In particular, let $\delta \lambda\!=\!\varepsilon\!-\!i\eta$ with both $\varepsilon$ and $\eta$ real and positive. We have
\begin{equation}
    \lambda_m + \delta \lambda = \mathcal{F}(g_m + \delta g) \approx \mathcal{F}(g_m) + \mathcal{F}'(g_m) \delta g + \frac{1}{2} \mathcal{F}''(g_m) \delta g^2
\end{equation}
Since $g_m$ is an extremal point (maximum in this case) for $\mathcal{F}$, the first derivative is zero and we are left with
\begin{equation}
    \delta \lambda \approx \frac{1}{2} \mathcal{F}''(g_m) \delta g^2 \quad \implies \quad \delta g \approx \sqrt{\frac{2(z - \lambda_m)}{\mathcal{F}''(g_m)}}
\end{equation}
Note that $\mathcal{F}''(g_m)\!<\!0$ as we previously pointed out, so $\delta g$ is purely imaginary, as expected. From expression \eqref{eq:spectrum} we finally have that
\begin{equation}
    \rho_{\mathcal{M}} (\lambda) \sim \sqrt{\lambda - \lambda_m}
\end{equation}
The procedure to obtain the spectrum near the other edge, $\lambda_M$, is identical
\begin{equation}
    \rho_{\mathcal{M}} (\lambda) \sim \sqrt{\lambda_M - \lambda}
\end{equation}
We denote this as GOE-like spectrum, since it is completely dominated by the matrix of the Gaussian ensemble.

\subsubsection*{Case 2: DIAG-like spectrum}

If the maximum does not exist ($\Lambda\!>\!0$) we can still expand around $(g_m\!=\!- \frac{a_m}{J^2}, \lambda_m\!=\!\mathcal{F} \left( g_m \right) )$, where $g_m$ is now at the edge of the domain. This time, though, the first derivative is not zero. In particular we have
\begin{equation}
    \mathcal{F}'(g_m) = - J^2 \left[ 1 - \int_{a_m}^{a_M} da \, P(a) \frac{J^2}{(a - a_m)^2} \right]
\end{equation}
Let us momentarily assume that the distribution of the diagonal elements $P(a)$ follows a power law at small $a$, $P(a)\!=\!\mathcal{A}_d (a\!-\!a_m)^\nu$, with $\nu\!=\!\frac{3}{2}\!+\!\alpha$ (we will prove this in a later section, Appendix \ref{app:curv_distribution}). Note that
\begin{eqnarray}
    \mathcal{F}'(g_m) \propto \int_{a_m}^{a_M} da \, (a - a_m)^{\nu-2} \\
    \mathcal{F}''(g_m) \propto \int_{a_m}^{a_M} da \, (a - a_m)^{\nu-3}
\end{eqnarray}
so we need $\nu\!>\!1$ for the convergence of the first derivative and $\nu\!>\!2$ for the second derivative. In general, for the convergence of the $n^{th}$ derivative, we need $\nu\!>\!n$, hence $\alpha\!>\!n\!-\!\frac{3}{2}$. In Ref.~\cite{bouchbinder_low-frequency_2021} the shape of the spectrum was worked out for the case $\alpha\!=\!0$ ($\nu\!=\!\frac{3}{2}$). In that case, the integral in the second derivative (and all the ones in higher order derivatives) did not converge and the power expansion of $\mathcal{F}$ had to be interrupted after the first order
\begin{equation}\label{eq:expansion_case2}
    \lambda_m + \delta \lambda =\mathcal{F}(g_m + \delta g) = \mathcal{F}(g_m) + \mathcal{F}'(g_m)\delta g + \delta \mathcal{F} (\delta g) \quad ,
\end{equation}
where $\delta \mathcal{F}$ contains all the non-analytical contributions coming from higher-order non-converging derivatives. In the limit of small perturbation, $\delta g\!\to\!0$, it was possible to obtain a simple form of the non-analytical contribution $\delta \mathcal{F}(\delta g)$, which could then be plugged back in \eqref{eq:expansion_case2} and inverted perturbatively to obtain $\delta g (\lambda\!-\!\lambda_m)$.

When $\alpha\!>\!0$ higher-order derivatives progressively start to converge, and additional contributions need to be considered in the expansion \eqref{eq:expansion_case2}. Following the same procedure just sketched for the case $\alpha\!=\!0$, it is possible to obtain the shape of the spectrum near the edge for any value of the exponent. We will provide here the derivation for the case $\alpha\!=\!1$ ($\nu\!=\!\frac{5}{2}$) and spell out the general result for arbitrary exponent. For $\alpha\!=\!1$, the second derivative $\mathcal{F}''(g_m)$ converges but $\mathcal{F}'''(g_m)$ does not, so Eq.~\eqref{eq:expansion_case2} now reads
\begin{equation}\label{eq:expansion_case2_new}
    \lambda_m + \delta \lambda =\mathcal{F}(g_m + \delta g) = \mathcal{F}(g_m) + \mathcal{F}'(g_m)\delta g + \frac{1}{2}\mathcal{F}''(g_m)\delta g^2 + \delta \mathcal{F} (\delta g)
\end{equation}
where, again, the non-analyticity of the function is all contained in $\delta\mathcal{F}$ (in general, for $ n\!<\!\nu\!<\!n\!+\!1$, which in terms of the exponent of the stiffness distribution, corresponds to $n\!-\!\frac{3}{2}\!<\!\alpha\!<\!n\!-\!\frac{1}{2}$, we need to expand $\mathcal{F}(g_m)$ up to its $n^{th}$ derivative). The non-analytic term in the expansion reads
\begin{equation}
\begin{split}
    \delta \mathcal{F} (\delta g) &= \mathcal{F}(g_m + \delta g) - \mathcal{F}(g_m) - \mathcal{F}'(g_m)\delta g - \frac{1}{2} \mathcal{F}''(g_m) \delta g^2 \\
    &= J^8 \delta g^3 \int_{a_m}^{a_M} da \, P(a) \frac{1}{(a - a_m + J^2 \delta g)(a - a_m)^3} \quad .
\end{split}
\end{equation}
We perform the following substitution and change of variable
\begin{equation}\label{eq:change_variable}
    q(a) = \frac{P(a)}{(a - a_m)^\nu} \quad , \quad x = \frac{a - a_m}{\delta g}
\end{equation}
With these changes the non-analytic term in the expansion reads
\begin{equation}
    \delta \mathcal{F} (\delta g) =  J^8 \delta g^\nu \int_0^{\frac{a_M - a_m}{\delta g}} dx \, \frac{q(x \delta g + a_m) x^{\nu - 3}}{x + J^2}
\end{equation}
Note that $\lim_{a \to a_m} q(a)\!=\!\mathcal{A}_d$, the prefactor of the distribution of diagonal terms, so that in the limit of a small perturbation $\delta g$ we have
\begin{equation}
    \delta \mathcal{F} (\delta g) \xrightarrow{\delta g\rightarrow 0} J^8 \delta g^\nu \mathcal{A}_d \int_0^\infty dx \, \frac{x^{\nu - 3}}{x + J^2} = J^7 \pi \delta g^{5/2} \mathcal{A}_d
\end{equation}
where in the last equality we used that $\alpha\!=\!1$ ($\nu\!=\!\frac{5}{2}$) so that the integral evaluates to $\frac{\pi}{J}$. With the simple expression for the non-analytic part in our hands, we can rewrite the expansion \eqref{eq:expansion_case2_new}
\begin{equation}\label{app:eq:perturbation_final}
    \delta \lambda \approx -J^2 \Lambda \delta g - \frac{1}{2} J^6 \Gamma \delta g^2 + J^7 \pi \delta g^{5/2} \mathcal{A}_d
\end{equation}
where we used the definition of $\Lambda$, Eq.~\eqref{app:eq:Lambda_def}, for the first derivative and we additionally defined $\mathcal{F}''(g_m)\!=\!- J^6 \Gamma$ (note that both $\Lambda$ and $\Gamma$ are finite and positive). We can invert perturbatively Eq.~\eqref{app:eq:perturbation_final} and obtain
\begin{equation}
    \delta g = - \frac{1}{J^2 \Lambda} \delta \lambda - \frac{\Gamma}{2 \Lambda^3} \delta \lambda^2 +\frac{\pi \mathcal{A}_d}{\Lambda^{7/2}} (-\delta \lambda)^{\frac{5}{2}}
\end{equation}
Finally, taking the imaginary part yields
\begin{equation}
    \rho_{\mathcal{M}} (\lambda) = \frac{\mathcal{A}_d}{\Lambda^{7/2}} (\lambda - \lambda_m)^{\frac{5}{2}}
\end{equation}

We can generalize this result to any value of $\alpha$, by making use of the finite geometric series
\begin{equation}
    \frac{1}{a- a_m +J^2 \delta g} = \sum_{k=0}^n \frac{(-1)^k(J^2\delta g)^k}{(a-a_m)^{k+1}} + \frac{(-1)^{n+1}(J^2\delta g)^{n+1}}{(a-a_m)^{n+1}(a-a_m+J^2\delta g)}
\end{equation}
to simplify the Taylor expansion of $\mathcal{F}(g_m\!+\!\delta g)$. For $\lfloor \nu \rfloor\!=\!n$ the non-analytic term reads
\begin{equation}
    \delta\mathcal{F}(\delta g) = (-1)^n J^{2n + 4} \delta g^{n+1} \int_{a_m}^{a_M} da \, \frac{P(a)}{(a-a_m)^{n+1}(a-a_m+J^2\delta g)} \quad .
\end{equation}
Using the same change of variable as in \eqref{eq:change_variable}, and in the limit of $\delta g\!\to\!0$, we obtain 
\begin{equation}
    \delta \mathcal{F} (\delta g) \approx \frac{(-1)^n J^{2\nu +2} \pi}{\sin[\pi(\nu - n)]} \mathcal{A}_d \delta g^{\nu} \quad .
\end{equation}
Again, we want to invert the expansion of $\delta \lambda\!=\!z\!-\!\lambda_m\!=\!\mathcal{F}(g_m\!+\!\delta g)\!-\!\lambda_m$ and isolate $\delta g$. The analytic terms in the expansion will produce only integer powers of $\delta \lambda$ in $\delta g (\delta \lambda)$, which are all real and do not contribute to the spectrum $\rho_\mathcal{M}(\lambda)$, given by \eqref{eq:spectrum}. The non-analytic term, instead, gives $\delta g\!\approx\!(-\delta \lambda)^\nu$ to leading order, from which we obtain that the spectrum scales as $\rho_{\mathcal{M}}(\lambda)\!\sim\!(\lambda\!-\!\lambda_m)^{\nu}\!=\!(\lambda\!-\!\lambda_m)^{\frac{3}{2}+\alpha}$.

\subsection{The replica symmetry and its breaking}\label{app:RS_RSB}
\subsubsection{The RS phase}
In this section we want to give a short overview of the replica framework used to solve the model, and in particular of its connection to the closure of the gap in the density of states, $\mathcal{D}(\omega)$. We refer to Ref.~\cite{bouchbinder_low-frequency_2021} for the details of the derivation.

We consider the replicated partition function of the KHGPS model at a finite temperature $T\!=\!1/ \beta$ averaged over the disorder, $\langle{\mathcal{Z}^n}\rangle$. Assuming a replica-symmetric ansatz for the overlap matrix, i.e. $Q_{ab}^{RS}\!=\!(\Tilde{q}\!-\!q)\delta_{ab}\!+\!q$, we have
\begin{equation}
    \langle{Z^n}\rangle = \int dq \, d\Tilde{q} \, e^{-NS[q,\Tilde{q}]}
\end{equation}
with the replica-symmetric action 
\begin{equation}
    S[q,\Tilde{q}] = \frac{(\beta J)^2}{4} n \left[\Tilde{q}^2 + (n-1) q^2 \right] - \log \left[ \int d \kappa \, P(\kappa) \int \mathcal{D}z \left( \int dx \, e^{ - \beta V_{\rm eff} } \right)^n \right]
\end{equation}
and the effective potential
\begin{equation}
    V_{\rm eff} = \frac{\kappa}{2} x^2 + \frac{x^4}{4!} - \frac{\beta J^2}{2} (\Tilde{q} - q) x^2 - (J \sqrt{q} z + h )x
\end{equation}
The auxiliary variable $z$ comes from a Hubbard--Stratonovich transformation and is normally distributed (we denote with $\mathcal{D}z$ the Gaussian integration measure).

In the thermodynamic limit, $N\!\to\!\infty$, we can apply the saddle point approximation. The values of $q$ and $\Tilde{q}$ for which the action assumes its minimum value are selected by a pair of self-consistent equations, which in the $n\!\to\!0$ limit become
\begin{gather}
    \Tilde{q} = \int d \kappa \, P(\kappa) \int \mathcal{D}z \langle x^2 \rangle_\beta \quad , \\
    q = \int d \kappa \, P(\kappa) \int \mathcal{D}z \langle x^2 - \frac{z x}{\sqrt{q}\beta J} \rangle_\beta \quad ,
\end{gather}
where the angular brackets denote an average over the effective potential:
\begin{equation}
    \langle \mathcal{O}(x) \rangle_\beta  \equiv \frac{\int dx \, \mathcal{O}(x) e^{-\beta V_{\rm eff}(x)} }{\int dx \, e^{-\beta V_{\rm eff}(x)} } \quad .
\end{equation}
It is also useful to define the quantity $\chi\!=\!\beta (\Tilde{q}\!-\!q)$. The free energy of the system in the thermodynamic limit, $\mathcal{F}\!=\!\langle \log \mathcal{Z}\rangle\!=\!\lim_{n\to 0} \frac{1}{n} \langle \mathcal{Z}^n \rangle$, reads
\begin{equation}
    \mathcal{F}[q, \tilde{q}] = \frac{(\beta J)^2}{4}(\tilde{q}^2 - q^2) - \int d\kappa \, p(\kappa) \int \mathcal{D}z \, \log \left( \int dx \, e^{-\beta V_{\rm eff}} \right)
\end{equation}

We consider, now, the low-temperature limit. We first notice that in the $\beta\!\to\!\infty$ limit we have $q\!\to\!\Tilde{q}$ linearly, so $\chi$ will tend to a finite value. Then, we also notice that in the low-temperature limit the averages $\langle\mathcal{O}(x)\rangle_\beta$ are dominated by the ground state $x^*$ of the effective potential $V_{\rm eff}$, which we can cast in this simplified form:
\begin{equation}\label{eq:effpotKHGPS}
    V_{\rm eff} = \frac{1}{4!} x^4 + \frac{m}{2} x^2 - H x
\end{equation}
where we defined
\begin{gather}
    m = \kappa - J^2\chi \quad , \label{app:eq:m_def} \\
    H =h + J z \sqrt{\tilde{q}} \quad .
\end{gather}
The effective field $H$ is normally distributed around $h$ with variance $J^2 \tilde{q}$, while the effective stiffness $m$ follows the same probability distribution as $\kappa$, but shifted by $J^2 \chi$. The correlation between different replicas is encoded in the two quantities $\tilde{q}$ and $\chi$, which in terms of the new effective stiffness distribution read
\begin{gather}\label{eq:self_const_eq_KHGPS}
    \tilde{q} = \int dm \, P(m) \int \mathcal{D}z \, \left( x^* \right)^2 = \big\langle (x^*)^2\big\rangle_{z,m} \quad , \\
    \chi = \int dm \, P(m) \int \mathcal{D}z \, \frac{z  x^*}{ J\sqrt{\tilde{q}}} = \int dm \, P(m) \int \mathcal{D}H \, \lim_{\beta \to \infty} \frac{d \langle x \rangle_\beta}{ dH} \quad . \label{eq:chi_finiteT}
\end{gather}
Note that the derivative and the low-temperature limit do not always commute: when the effective stiffness of the potential is negative, $V_{\rm eff}$ has two minima and $x^*$ is not a smooth function of $H$. The proper limit then reads
\begin{equation}
    \lim_{\beta\to \infty} \frac{d\langle x \rangle_\beta}{dH} = \frac{1}{V''_{\rm eff}[x^*(H,m)]} + [x^*(0^+,m)-x^*(0^-,m)]\delta(H)
\end{equation}
Whenever $m_m\!\ge\!0$, $\chi$ can be equivalently expressed as
\begin{equation}\label{app:eq:chi}
    \chi = \int dm \, P(m) \int \mathcal{D}z \, \frac{1}{ m +\frac{1}{2}(x^*)^2} = \int da\,P(a) \frac{1}{a - J^2\chi} = \big\langle \frac{1}{V''_{\rm eff}[x^*(z,m)]}\big\rangle_{z,m} \quad .
\end{equation}

Finally, the expression of the replicon eigenvalue of the Hessian of the replicated free energy \cite{nishimori_statistical_2001} reads
\begin{equation}\label{app:eq:replicon_div}
    \lambda_R =1- J^2\int dm \, P(m) \int \mathcal{D}H \, \lim_{\beta \to \infty} \left(\frac{d \langle x \rangle_\beta}{ dH}\right)^2 \quad .
\end{equation}
which, again, in the case $m_m\!\ge\!0$ is equivalent to
\begin{equation}\label{app:eq:replicon}
    \lambda_R =1- J^2\int dm \, P(m) \int \mathcal{D}z \, \frac{1}{ [m +\frac{1}{2}(x^*)^2]^2}  = 1- J^2 \int da\,P(a) \frac{1}{(a - J^2\chi)^2} \quad .
\end{equation}

\subsubsection{The transition to the RSB phase}\label{app:breakingRSB}

The value of the replicon mode determines the stability of the system and the solution spelled out in the previous section is valid only as long as $\lambda_R\!\ge\!0$. In this section we will consider how $\lambda_R$ can become negative and what the physical consequences for the system are.

\paragraph*{Case 1 -- Continuous transition:}
Let us assume first that the replicon is able to smoothly become negative in a region of the parameter space. From Eq.~\eqref{app:eq:replicon} we have that when $\lambda_R\!=\!0$ then
\begin{equation}\label{app:condition1}
    1 = J^2\int da \, P(a) \frac{1}{(a - J^2\chi)^2}
\end{equation}
For the GOE-like region, where $\mathcal{F}$, Eq.~\eqref{eq:f_function}, has a maximum strictly inside its domain ($g_m\!>\!-a_m/J^2$):
\begin{equation}
    \mathcal{F}'(g_m) = 0 \quad \text{for } g_m >- \frac{a_m}{J^2}  \quad \implies \quad 1 =  \int da \, P(a)  \frac{J^2}{(a + J^2g_m)^2} 
\end{equation}
condition \eqref{app:condition1} implies $\chi\!=\!-g_m$. If we compute now the lower bound of the spectrum in this regime, $\lambda_m\!=\!\mathcal{F}(g_m)$, we obtain that
\begin{equation}
    \lambda_m = J^2 \int da \, P(a) \left( \chi - \frac{1}{a - J^2 \chi} \right) = J^2\chi - J^2\int da \, P(a) \frac{1}{a - J^2 \chi} = 0
\end{equation}
where we recognized expression \eqref{app:eq:chi}. We conclude that when the replica symmetry breaks continuously, the spectrum in the GOE-like regime becomes gapless, $\rho_\mathcal{M}(\lambda)\!\sim\!\lambda^{1/2}$.

\paragraph*{Case 2 -- Discontinuous transition:} The picture just described above breaks down when double wells appear in the effective potential. The discontinuous jump in $x^*(H,m)$ makes the replicon, Eq.~\eqref{app:eq:replicon_div}, diverge to $-\infty$, breaking the replica symmetry. At the transition, when $m_m\!=\!0$, from the definition of the diagonal elements and of the effective stiffness, Eq.~\eqref{app:eq:m_def}, we have $a_m\!=\!\kappa_m\!=\!J^2 \chi$. We also note that it is possible to write
\begin{equation}
    \Lambda = 1 - J^2 \int da \, P(a) \frac{1}{(a-a_m)^2} =1 - J^2 \int da \, P(a) \frac{1}{(a-J^2\chi)^2 } = \lambda_R
\end{equation}
Given that in the RS phase the replicon must be positive, this discontinuous breaking can happen exclusively in the DIAG-like regime, in which $\Lambda\!>\!0$. There the edge of the spectrum $\lambda_m\!=\!\mathcal{F}(g_m)$ falls at the boundary of the domain, $g_m\!=\!-a_m/J^2$, which implies that
\begin{equation}
    \lambda_m = \mathcal{F}(-\frac{a_m}{J^2}) = \mathcal{F}(-\chi) = J^2\chi - J^2\int da \, P(a) \frac{1}{a-J^2\chi} = 0
\end{equation}
where we used the definition of $\chi$. Hence the spectrum is gapless, $\rho_\mathcal{M}(\lambda)\!\sim\!\lambda^\nu$.

\subsubsection{The phase diagram}

A special point $(J^*,h^*)$ separates the continuous from the discontinuous RSB transition, in which both $\lambda_R\!=\!0$ and $m_m\!=\!0$. This corresponds exactly to having
\begin{equation}
    \Lambda = \lambda_R = 1- J^2 \int_{J^2\chi}^{a_M} da\,P(a) \frac{1}{(a - J^2\chi)^2} = 0
\end{equation}
We can numerically compute the location of the special point by choosing a particular form of the stiffness distribution $P(\kappa)$, which we left unspecified up to this point. We now fix it to a generic distribution from the power-law family, which properly normalized reads
\begin{equation}
    P(\kappa) = \frac{\alpha + 1}{(\kappa_M - \kappa_m)^{\alpha + 1}} (\kappa - \kappa_m)^\alpha \quad .
\end{equation}
By varying the exponent $\alpha$, the location of the special point changes, tracing a line $(J^*(\alpha), h^*(\alpha), \alpha)$ in the parameter space. Imposing the additional condition $m_m\!=\!0$ on $\lambda_R\!=\!0$, Eq.~\eqref{app:eq:replicon}, gives us
\begin{equation}\label{app:eq:lambda0_ath0}
    1= J^2\int_0^{m_M} dm \, \frac{\alpha+1}{w^{\alpha+1}} m^\alpha\int \mathcal{D}(z) \frac{1}{(m +\frac{1}{2}(x^*)^2)^2}
\end{equation}
where we defined $w\!=\!\kappa_M\!-\!\kappa_m$ to ease the notation, and we remind the reader that $P_m(m)\!=\!P(m\!+\!J^2\chi)$. Instead, $\chi$ reads
\begin{equation}\label{app:eq:chi0_ath0}
    \chi = \int_0^{m_M} dm \, \frac{\alpha+1}{w^{\alpha+1}} m^\alpha \int \mathcal{D}z \frac{1}{m + \frac{1}{2}(x^*)^2}
\end{equation}

We want to know now for which exponent $\alpha\!=\!\alpha_0$ the line $(J^*, h^*, \alpha)$ reaches the $h\!=\!0$ axis. To do so let us impose the additional condition $h^*\!=\!0$, which centers the distribution of the Gaussian field $H$ around zero. This causes the collapse of the $\tilde{q}\!=\!\langle (x^*)^2\rangle$ integral to zero, and without any external field to break the $\mathbb{Z}_2$ symmetry, the above integrals, Eqs.~\eqref{app:eq:lambda0_ath0} and \eqref{app:eq:chi0_ath0}, simplify to, respectively,
\begin{equation}\label{app:eq:lambda_h0_simple}
    1= J^2\int_0^{m_M} dm \, \frac{\alpha+1}{w^{\alpha+1}} m^{\alpha-2} = J^2 \frac{\alpha + 1}{\alpha - 1} w^{-2}
\end{equation}
and
\begin{equation}\label{app:eq:chi_h0_simple}
    \chi = \int_0^{m_M} dm \, \frac{\alpha+1}{w^{\alpha+1}} m^{\alpha -1} = \frac{\alpha + 1}{\alpha} w^{-1}
\end{equation}
The value of $\alpha_0$ is then obtained by comparing the two equations, together with the additional relation $\kappa_m\!=\!J^2\chi$ coming from imposing $m_m\!=\!0$. We obtain
\begin{equation}
    \alpha_0 = \frac{w}{w - \kappa_m}
\end{equation}
and also
\begin{equation}
    J^*(\alpha_0) = \sqrt{\frac{\alpha_0}{\alpha_0 + 1} \kappa_m w} \quad.
\end{equation}
For the system simulated in the present paper, we used $\kappa_m\!=\!0.1$ and $\kappa_M\!=\!1.0$, yielding $\alpha_0\!=\!1.125$. The result obtained here tells us that when $\alpha$ increases, the portion of the parameter space in which the continuous RSB transition takes place shrinks until the special point separating it from the discontinuous branch of the transition line touches the zero field plane, at $(J^*(\alpha_0), 0 , \alpha_0)$, see Fig.~\ref{fig:phase_diagram_3d}. From this point on, the only admitted RSB transition is of the discontinuous type, with a density of states governed by the distribution of diagonal elements, $P(a)$. 

Finally, we note that the RS solution must still be stable when $m_m$ reaches zero, therefore we must have a non-negative replicon, $\lambda_R\!\ge\!0$. This turns Eq.~\eqref{app:eq:lambda_h0_simple} into an inequality
\begin{equation}
    1\ge J^2 \frac{\alpha + 1}{\alpha - 1} w^{-2}
\end{equation}
which, when combined with Eq.~\eqref{app:eq:chi_h0_simple} and inverted in favor of $\alpha$, yields $\alpha\!\ge\!\alpha_0$, a condition that can only be met if $\kappa_M\!>\!2\kappa_m$. This means that in order to have a discontinuous RSB transition on the $h\!=\!0$ axis, the support of the stiffness distribution needs to be large enough.

%%%%%%%%%%%%%%%%%%%%%%%%%%
\begin{figure}
\centering
{\includegraphics[clip, trim=2cm 0cm 0cm 0cm,scale=0.45]{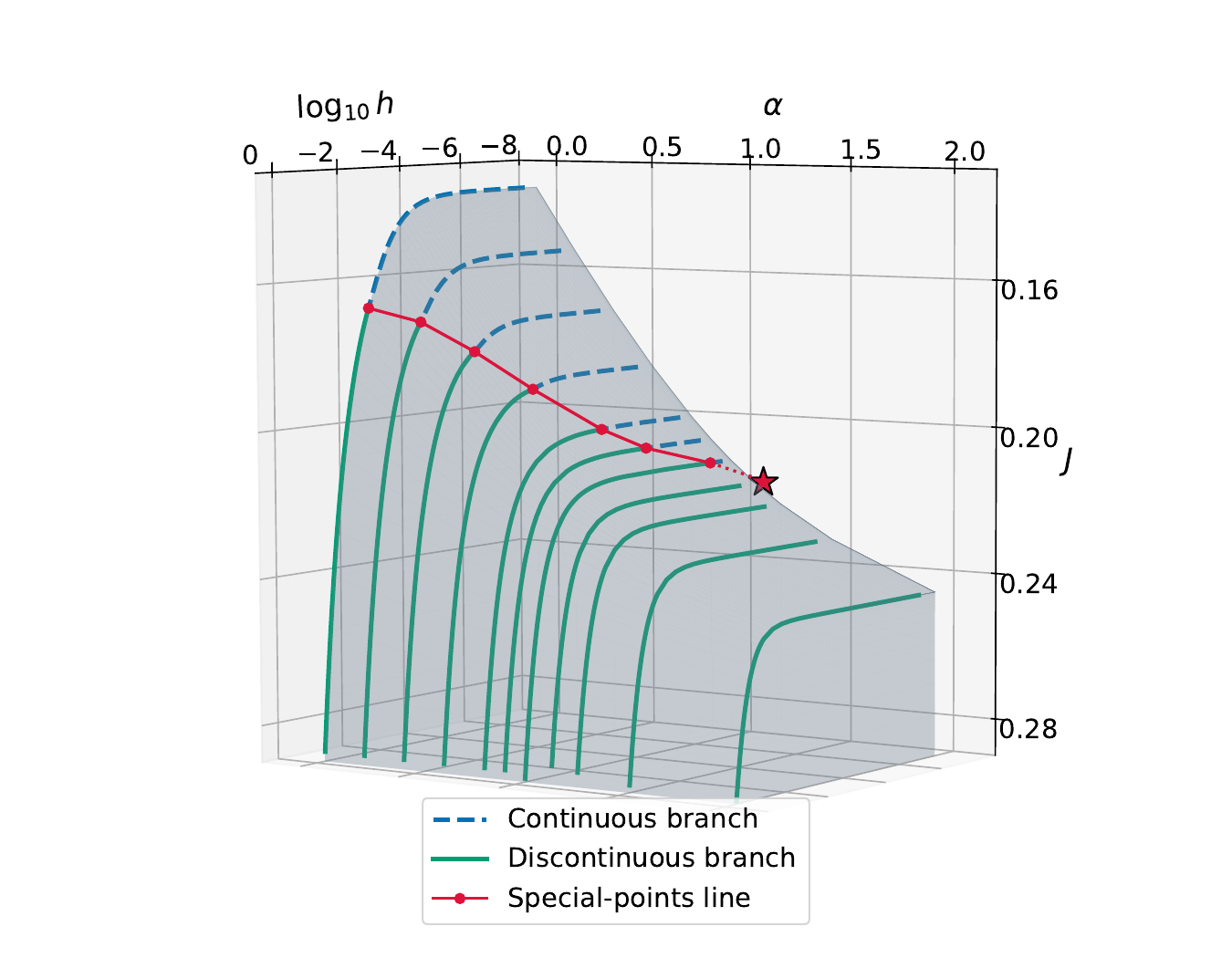}}
\caption{Phase diagram of the KHGPS model in the $(J, h, \alpha)$ parameter space. The shaded volume represents the phase with broken replica symmetry. The dashed blue lines represent the continuous RSB transition, with $\mathcal{D}(\omega)\!\sim\!\omega^2$, while the solid green lines indicate the discontinuous transition, with the $\alpha$-dependent density of states. The red line connects the special points $(J^*(\alpha), h^*(\alpha))$ separating the two regimes. When we increase $\alpha$, the special points move to smaller and smaller values of the external field $h$, and eventually touch the $h\!=\!0$ plane. For the considered parameters $\kappa_m\!=\!0.1$, $\kappa_M\!=\!1.0$, this happens at $\alpha_0\!=\!1.125$, $J^*(\alpha_0)\!=\!0.2183$, marked by the red star (which we drew at a low, but still finite, value of $h$, given the logarithmic axis). It is visible from the plot that no $\omega^2$ branch exists above those values.}
\label{fig:phase_diagram_3d}
\end{figure}
%%%%%%%%%%%%%%%%%%%%%%%%%%

\subsection{Curvature distribution}\label{app:curv_distribution}

To compute $P(\tilde{a})$ we first need to study the ground state of the effective potential, $x^*$, which is the solution of the following depressed cubic equation
\begin{equation}\label{app:eq:depressed_cubic}
	x^3 + 6 m x - 6 H = 0
\end{equation}
whose discriminant is
\begin{equation}
	\Delta \propto m^3 + \frac{9}{8} H^2 \, .
\end{equation}
Whenever $\Delta\!>\!0$, Eq.~\eqref{app:eq:depressed_cubic} yields a single solution, while the equation has three real roots when $\Delta\!<\!0$, and we need to select the one corresponding to the global minimum (additionally, note that $\Delta\!<\!0$ implies $9 H^2\!<\!- (2m)^3$, which in turn implies $m\!<\!0$). In both cases we can obtain an explicit form for $x^*(m,H)$ using the known expressions for the roots of cubic equations
\begin{equation}
	x^*(m, H) = 
	\begin{cases}
		2 \sgn (H) \sqrt{2 |m|} \sinh \left[ \frac{1}{3} \arcsinh \left(\frac{3|H|}{(2|m|)^{3/2}} \right) \right] \quad & \text{if } m > 0 \\
        2 \sgn (H) \sqrt{2 |m|} \cosh \left[ \frac{1}{3} \arccosh\left(\frac{3|H|}{(2|m|)^{3/2}}\right) \right] \quad &\text{if } m < 0 \text{ }\land \text{ } \Delta > 0 \\
        2 \sgn (H) \sqrt{2 |m|} \cos \left[ \frac{1}{3} \arccos \left(\frac{3|H|}{(2|m|)^{3/2}} \right) \right] \quad &\text{if } m < 0 \text{ }\land \text{ } \Delta < 0
	\end{cases}
\end{equation}

%%%%%%%%%%%%%%%%%%%%%%%%%%
\begin{figure}
\centering
{\includegraphics[scale=0.63]{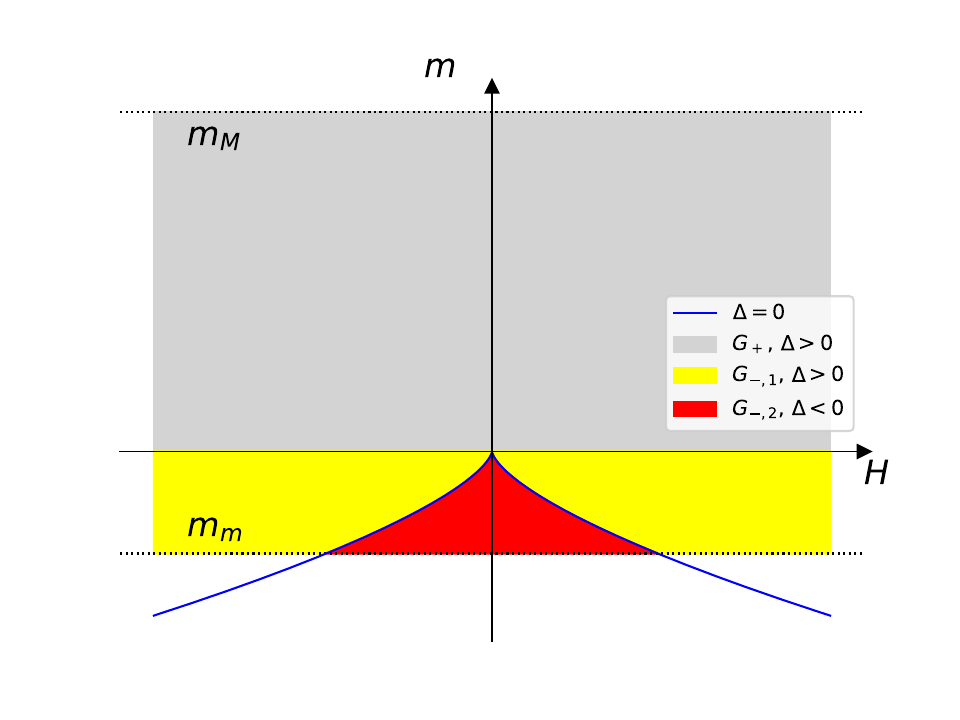}}
\caption{Integration domain of the cumulative probability distribution $G(\tilde{a})$. The domain is divided into three regions, in each of which a different expression for the global minimum $x^*$ holds.}
\label{fig:integr_domain}
\end{figure}
%%%%%%%%%%%%%%%%%%%%%%%%%%

With the expression of $x^*$ in hand we can now obtain the distribution of curvature $P(\tilde{a})$. We proceed via the cumulative:
\begin{equation}
	G(\tilde{a}) = \int_{0}^{\tilde{a}} d \tilde{a}' \, P(\tilde{a}') = \int dm \,  P_m(m) \int dH \, P_H(H)  \Theta \left[ m + \frac{1}{2} (x^{*})^2 < \tilde{a} \right]
\end{equation}
where we added a subscript to the two probability distributions, $P_X(X),$ to distinguish them easily and avoid confusion at later stages. Given the different expressions of $x^*$ for different values of $m$ and $H$, we need to split the integration domain in different portions (see Fig.~\ref{fig:integr_domain}). We consider here the region $G_+$, where $m\!>\!0$ (the others, $G_{-,1}$ and $G_{-,2}$, can be obtained by following the same procedure)
\begin{equation}
	G_+(\tilde{a}) = \int_{0}^{\tilde{a}}   dm \, P_m(m) \int_{- \infty}^{\infty} dH \, P_H(H) \Theta \left[ m + \frac{1}{2} (x^{*})^2 < \tilde{a} \right]
\end{equation}
We perform the following changes of variable
\begin{equation}
	\xi = \frac{m}{\tilde{a}} \quad , \quad \eta = \frac{H}{(\xi \tilde{a})^{3/2}}
\end{equation}
The determinant of the Jacobian of the transformation is $|J|\!=\!\tilde{a}^{5/2} \xi^{3/2}$ and the integral now reads
\begin{equation}\label{app:eq:midresult}
	G_+(\tilde{a}) = \int_{0}^{1}  d \xi \, (\tilde{a}^{5/2} \xi^{3/2}) P_m(\xi \tilde{a}) \int_{- \infty}^{\infty}  d \eta \, P_H\left(\eta (\xi \tilde{a})^{3/2}\right) \Theta \left[ \xi \left(1 + \frac{1}{2} y^2 \right) < 1 \right]
\end{equation}
where we defined $y\!=\!\frac{x^*}{\sqrt{\xi \tilde{a}}}$, which in the case under current study ($m\!>\!0$) reads
\begin{equation}\label{ypsilon}
	y = 2\sqrt{2} \sinh \left[ \frac{1}{3} \arcsinh \left( \frac{3 \eta}{2 \sqrt{2}} \right) \right]
\end{equation}

The Heaviside function in Eq.~\eqref{app:eq:midresult} crops the integration over the whole real axis to a finite region. The integral is still hard to solve exactly, but we can obtain its approximate scaling. As we will show, the scaling depends also on the external field $h$ and for this reason it is convenient to separate the discussion in two cases.

\subsubsection{Finite external field}

Let us first consider the case where the external field $h$ is finite. When the width $\sigma\!=\!J\sqrt{\tilde{q}}$ of the distribution $P_H$ of the effective field is large with respect to the integration interval, we can use the auxiliary formula 
\begin{equation}\label{app:eq:auxiliary_formula}
	\int_{-\varepsilon}^{\varepsilon}  dH \, P_H(H) = \frac{1}{2} \left[ \erf \left( \frac{h + \varepsilon}{J \sqrt{2 \tilde{q}}} \right) - 
    \erf \left( \frac{h - \varepsilon}{J \sqrt{2 \tilde{q}}} \right) \right] \approx  \frac{2 \varepsilon}{ \sqrt{2 \pi J^2 \tilde{q}}} e^{-\frac{h^2}{2 J^2 \tilde{q}}}  = P_H(0) \int_{-\varepsilon}^{\varepsilon} d H
\end{equation}
to approximate the Gaussian distribution with its value at the center of the interval, $P_H(0)\!=\!\frac{1}{ \sqrt{2 \pi J^2 \tilde{q}}} e^{-\frac{h^2}{2 J^2 \tilde{q}}}$. The approximation in \eqref{app:eq:auxiliary_formula} is valid only when the higher order terms in the expansion of $P_H(H)$ near zero are negligible with respect to the zeroth-order term. This condition can formally be expressed as
\begin{equation}
    \frac{\varepsilon^2}{6}\left(\frac{P''_H(0)}{P_H(0)} \right) = \frac{\varepsilon^2}{6\sigma^2}\left(\frac{h^2}{\sigma^2}-1\right) \ll 1 \quad ,
\end{equation}
or equivalently as $\varepsilon\!\ll\!\min\left(\sigma, \sigma^2 / |h|\right)$. Considering that the $H$ field integration window scales like $\tilde{a}^\frac{3}{2}$ (the variables $\eta$ and $\xi$ are just $\mathcal{O}(1)$ terms) and that $\sigma^2\!\sim\!\tilde{q}\!\sim\!h^2$ at small external field, the approximation holds in the limit $\tilde{a}\!\ll\!\tilde{a}_X$, with the upper bound that scales like $\tilde{a}_X\!\sim\!h^{2/3}$.

In the limit of small curvature $\tilde{a}$ we can then rewrite $G_+$ as
\begin{equation}
    G_+ (\tilde{a}) = \tilde{a}^{5/2} P_H(0) \int_{0}^{1}  d \xi \,  \xi^{3/2} P_m(\xi \tilde{a}) \int  d \eta \, \Theta \left[ \xi < \left(1 + \frac{1}{2} y^2 \right)^{-1} \right] \quad .
\end{equation}
We remind the reader that
\begin{equation}
    P_m(m) = \frac{\alpha + 1}{(\kappa_M - \kappa_m)^{\alpha + 1}} (m - m_m)^\alpha \quad ,
\end{equation}
and we substitute it in $P_m (\xi \tilde{a})$. Additionally bounding the integration interval through the Heaviside function, the integral becomes
\begin{equation}
    G_+ (\tilde{a}) = \frac{(\alpha + 1) P_H(0)}{(\kappa_M - \kappa_m)^{\alpha + 1}} \tilde{a}^{5/2} \int  d \eta \, \int_{0}^{\left(1 + \frac{1}{2} y^2 \right)^{-1}}  d \xi \,  \xi^{3/2} (\xi \tilde{a} - m_m)^\alpha
\end{equation}
The two integrals are now solvable, starting from the inner one. If $m_m\!=\!0$ (equivalently, when we are on the transition line between the RS and the RSB phase) then $G_+$ reads
\begin{equation}\label{G_plus_on_line}
    G_+ (\tilde{a}) = \frac{(\alpha + 1)P_H(0)}{(\kappa_M - \kappa_m)^{\alpha + 1}} \tilde{a}^{\frac{5}{2}+ \alpha} \frac{2}{5 + 2\alpha} \int  d \eta \, \left(1 + \frac{1}{2} y^2 \right)^{-\left( \frac{5}{2} + \alpha \right)} \sim \tilde{a}^{\frac{5}{2}+ \alpha}
\end{equation}
while in the case $m_m\!<\!0$ the cumulative distribution is dominated by
\begin{equation}\label{G_plus_inside_line}
    G_+ (\tilde{a}) = \frac{(\alpha + 1)P_H(0)}{(\kappa_M - \kappa_m)^{\alpha + 1}} (-m_m)^{\alpha} \tilde{a}^{\frac{5}{2}} \frac{2}{5} \int  d \eta \, \left(1 + \frac{1}{2} y^2 \right)^{-\frac{5}{2}} \sim \tilde{a}^{\frac{5}{2}}
\end{equation}
at low values of $\tilde{a}$, and then crosses over to \eqref{G_plus_on_line} when $\tilde{a}$ increases. Finally, the outer integral in the variable $\eta$ can be easily solved numerically, remembering the expression for $y$, Eq.~\eqref{ypsilon}, and it just amounts to a numerical prefactor.

The same procedure holds for $G_{-,1}$ and for $G_{-,2}$, with three important differences: (i) we need to change the boundary of the integration region; (ii) we need to choose the correct expression for $x^*$, and hence for $y$; (iii) we modify the change of variable to $\xi\!=\!- \frac{m}{\tilde{a}}$. This yields the following results:
\begin{eqnarray}
    G_{-,1} (\tilde{a}) = \frac{(\alpha + 1) P_H(0)}{(\kappa_M - \kappa_m)^{\alpha + 1}} \tilde{a}^{5/2} \int_{\eta^2 > \frac{8}{9}}  d \eta \, \int_{0}^{\left(-1 + \frac{1}{2} y^2 \right)^{-1}}  d \xi \,  \xi^{3/2} (-\xi \tilde{a} - m_m)^\alpha \\
    G_{-,2} (\tilde{a}) = \frac{(\alpha + 1) P_H(0)}{(\kappa_M - \kappa_m)^{\alpha + 1}} \tilde{a}^{5/2} \int_{\eta^2 < \frac{8}{9}}  d \eta \, \int_{0}^{\left(-1 + \frac{1}{2} y^2 \right)^{-1}}  d \xi \,  \xi^{3/2} (-\xi \tilde{a} - m_m)^\alpha
\end{eqnarray}

Of course, $G_{-,1}$ and $G_{-,2}$ contribute only if $m_m\!<\!0$ so they always show the crossover from $\sim \tilde{a}^{\frac{5}{2}}$ to $\sim \tilde{a}^{\frac{5}{2} + \alpha}$. Summing the different contributions, we will have
\begin{equation}
    G(\tilde{a}) = (A_{G_+} + A_{G_{-,1}} + A_{G_{-,2}} ) \frac{e^{-\frac{h^2}{2 J^2 \tilde{q}}}}{ \sqrt{2 \pi J^2 \tilde{q}}}  \tilde{a}^{\frac{5}{2} + \alpha} \quad \implies \quad P(\tilde{a}) = \mathcal{A}_d \tilde{a}^{\frac{3}{2} + \alpha}
\end{equation}
where we defined the numerical prefactor coming from the integration region $R$ as $A_R$, $P_H(0)$ excluded, and we denote with the calligraphic style $\mathcal{A}_d$ the prefactor of the probability distribution.

Summarizing, we obtained that in the approximation of small enough $\tilde{a}$ with respect to the external field, $\tilde{a}\!\ll\!\tilde{a}_X\!\sim\!h^{2/3}$, the probability distribution of the curvature $\tilde{a}$ of the effective potential follows a power-law distribution with exponent $\nu\!=\!\frac{3}{2}\!+\!\alpha$. When the lower bound of the effective stiffness $m_m$ is negative, the distribution $P(\tilde{a})$ shows a crossover to an $\alpha$-independent distribution $P(\tilde{a})\!\sim\!\tilde{a}^{3/2}$ below a certain value $\tilde{a}_{\text{cross}}$.

\subsubsection{Small external field}

When the external field $h$ decreases towards zero, the region of curvature $\tilde{a}\!\ll\!\tilde{a}_X$ below which the approximation \eqref{app:eq:auxiliary_formula} holds shrinks, until eventually, in the limit $h\!\to\!0$, it disappears altogether. We need to find the scaling of $P(\tilde{a})$ above $\tilde{a}_X$. Considering the effective potential, Eq.~\eqref{eq:effpotKHGPS}, and the definition of $\tilde{q}$ as mean squared displacement of the oscillators, Eq.~\eqref{eq:self_const_eq_KHGPS}, when the field $h$ decreases also the minimum $x^*$ diminishes, scaling like $x^*\!\sim\!m^{-1} H$. For this reason, the integration region for $\eta$ in Eq.~\eqref{app:eq:midresult} becomes barely affected by the Heaviside function and the scaling of $P(\tilde{a})$ is dominated by $P_m(\tilde{a})$. In the limit $h\!\to\!0$ this becomes exact: we have that the mean squared displacement $\tilde{q}\!=\!0$, hence $x^*\!=\!0$ and the $P_H(H)$ distribution can be integrated out, leaving
\begin{equation}
    P(\tilde{a})= P_m(\tilde{a}) \sim \tilde{a}^\alpha \quad .
\end{equation}
We focused for simplicity on the case $m_m\!=\!0$. For $m_m\!<\!0$ the double wells keep $\tilde{q}\!>\!0$ even at $h\!=\!0$, so the $\tilde{a}^{3/2}$ law of the previous section still applies at low $\tilde{a}$.

These different scaling regimes of the effective curvature distribution become relevant when the system sits in the DIAG-like region of the parameter space. There, the spectrum follows $\rho_{\mathcal{M}}(\lambda)\!\sim\!(\lambda\!-\!\lambda_m)^\nu$, with the same scaling exponent $\nu$ as the distribution of the (actual) curvatures $a$ (same as $P(\tilde{a})$, given that the two are related by a simple constant shift of the argument, $a\!=\!\tilde{a}\!+\!J^2 \chi $). Hence, the density of states of the KHGPS system on the discontinuous (in the sense of the replicon $\lambda_R$) branch of the RSB transition scales as
\begin{equation}
    \mathcal{D}(\omega) \sim \begin{cases}
        \omega^{4+2\alpha} \quad &\text{ if $\omega\!<\!\omega_X$} \\
        \omega^{1+2\alpha} \quad &\text{ if $\omega_X\!<\!\omega\!<\!\sqrt{\kappa_M}$} 
    \end{cases}
\end{equation}
with $\omega_X\!\sim\!h^{1/3}$.

\section{The simplified KHGPS and its density of states}

\subsection{The spectrum of the Hessian}\label{App:spectrum_hessian}

To determine the  density of eigenvalues $\rho_\mathcal{M}(\lambda)$ of the Hessian matrix $\mathcal{M}_{ij}\!=\!-\frac{J}{N}\!+\!\delta_{ij}(\kappa_i\!+\!\frac{A}{2} x_i^2)$ notice first that $\mathcal{M}$ can be decomposed into a constant matrix $\mathcal{C}$ with elements $-\frac{J}{N}$ and a diagonal matrix $\mathcal{B}$ with elements $b_i\!=\!\kappa_i\!+\!\frac{A}{2} x_i^2$. The distribution of eigenvalues of the constant matrix is given by
\begin{equation}
    \rho_\mathcal{C}(\lambda)= \frac{(N-1)}{N}\delta(\lambda)+\frac{1}{N}\delta(\lambda+J)
\end{equation}
while the spectrum of the diagonal matrix is given by the distribution of the diagonal elements, i.e. $\rho_\mathcal{B}(\lambda)\!=\!\rho_\mathcal{B}(b_i =\lambda)$. In the thermodynamic limit $ \rho_\mathcal{C}(\lambda)\!=\!\delta(\lambda)$ and the distribution of eigenvalues of the $\mathcal{M}$-matrix is completely determined by the diagonal part.

We first obtain a useful expression for $\rho_\mathcal{B}(b_i)$ in terms of the joint probability distribution ${\mathcal{P}_i}(x,\kappa)$. To do so we consider that $b_i\!=\!\kappa_i\!+\!\frac{A}{2}x_i^2$ and we perform the substitution $y\!=\!\frac{A}{2} x_i^2$. We can then evaluate the joint probability of the stiffness $\kappa_i$ with the newly defined variable $y$, $\mathcal{Q}(y, \kappa_i)$
\begin{eqnarray}
    \mathcal{Q}(y,\kappa_i)=\frac{1}{\sqrt{2 A y}} \left(\mathcal{P}_i\left(\sqrt{\frac{2}{A}y},\kappa_i\right)+ \mathcal{P}_i\left(-\sqrt{\frac{2}{A} y},\kappa_i\right)\right)
\end{eqnarray}
Note that the support of the distribution for the variable $y$ is now from $0$ to $\infty$, while it stays the same for $\kappa_i$, i.e. $\kappa_i\!\in\![\kappa_{m}, \kappa_{M}]$. A second change of variable $b_i\!=\!\kappa_i\!+\!y$ leads to 
\begin{eqnarray}
    \mathcal{R}(b_i,\kappa_i)= \mathcal{Q}(b_i-\kappa_i,\kappa_i) =\frac{1}{\sqrt{2 A (b_i-\kappa_i)}} \left(\mathcal{P}_i\left(\sqrt{\frac{2}{A}(b_i-\kappa_i)},\kappa_i\right)+ \mathcal{P}_i\left(-\sqrt{\frac{2}{A} (b_i-\kappa_i)},\kappa_i\right)\right)
\end{eqnarray}
Finally, to obtain $\rho_\mathcal{B}(b_i)$ we just need to marginalize $\mathcal{R}(b_i,\kappa_i)$ with respect to $\kappa_i$. The integration region is bounded below by $\kappa_{m}$, and considering that $b_i\!=\!\kappa_i\!+\!y$, with $y\!>\!0$, we have that the upper bound is $\min(b_i,\kappa_{M})$. The probability distribution of the diagonal elements of the Hessian reads
\begin{equation}\label{app:spectrum_expr}
    \rho_\mathcal{B}(b_i)=\int_{\kappa_{m}}^{\min(b_i,\kappa_{M})} \frac{1}{\sqrt{2 A (b_i-\kappa)}} \left(\mathcal{P}\left(\sqrt{\frac{2}{A} (b_i-\kappa)},\kappa\right)+ \mathcal{P}\left(-\sqrt{\frac{2}{A} (b_i-\kappa)},\kappa\right)\right)d\kappa
\end{equation}
Given the symmetry of the Hamiltonian under the permutation of indices, the index $i$ can be dropped and we have that $\rho_\mathcal{B}(b_i)\!=\!\rho_\mathcal{B}(\lambda)$ for all the oscillators.

\subsection{The joint probability}\label{App:joint_prob}

The joint probability $\mathcal{P}(\tilde{x},\tilde{\kappa})\!=\!\mathcal{P}_i(\tilde{x},\tilde{\kappa})$ represents the probability of finding the $i$-th oscillator in position $\tilde{x}$ and with a stiffness $\tilde{\kappa}$, and it is identical for all the oscillators, given the already mentioned symmetry of the system under the permutation of indices. Alternatively, it can be obtained also from the joint probability distribution of the random fields $\{h_j\}$, the stiffnesses $\{\kappa_j\}$ and the positions $\{x_j\}$ of the spins (whose probability is just given by the Boltzmann weight), marginalized with respect to everything except the position and the stiffness of the $i$-th oscillator
\begin{equation}\label{app:eq:marginalizationP}
    \mathcal{P}_i(\tilde{x},\tilde{\kappa})= P(\tilde{\kappa})\int \Pi_{j\neq i}(dx_j d\kappa_j dh_j P(\kappa_j)P(h_j)) P(h_i)  dh_i  \frac{1}{\mathcal{Z}}e^{-\beta \mathcal{H}^S[x,x_i=\tilde{x}, \kappa_i=\tilde{\kappa}]}
\end{equation}
where $\mathcal{Z}$ is the partition function of the system. We can use this alternative definition to obtain a more useful expression for $\mathcal{P}(\tilde{x},\tilde{\kappa})$. To do that we add an auxiliary field $g$ such that the Hamiltonian reads
\begin{equation}
    \mathcal{H}^S= \sum_{i} \left[\frac{1}{2}\kappa_i x_i^2+ \frac{A}{4!} x_i^4-(h+h_i) x_i + g(x_i,\kappa_i)\right]-\sum_{i,j} \frac{J}{2N} x_i x_j \quad .
\end{equation}
Computing the functional derivative with respect to the auxiliary field $g(\tilde{x},\tilde{\kappa})$ of the logarithm of the partition function averaged over the quenched disorder we have
\begin{equation}
    \frac{\delta }{\delta g(\tilde{x},\tilde{\kappa})} \langle \log(\mathcal{Z})\rangle = - \int \Pi_{j}(dx_j d\kappa_j dh_j P(\kappa_j)P(h_j))  \int  \frac{\beta}{\mathcal{Z}}e^{-\beta \mathcal{H}^S[x]}  \sum_i \frac{\delta  g(x_i,\kappa_i) }{\delta g(\tilde{x},\tilde{\kappa})} = -\beta \sum_i \mathcal{P}_i(\tilde{x},\tilde{\kappa})
\end{equation}
where we have used  $\frac{\delta  g(x_i,\kappa_i) }{\delta g(\tilde{x},\tilde{\kappa})}\!=\!\delta(x_i\!-\!\tilde{x})\delta(\kappa_i\!-\!\tilde{\kappa})$ and where we recognized expression \eqref{app:eq:marginalizationP} from above. Since the joint probabilities are all identical, irrespective of the index, we obtain the desired expression
\begin{equation}\label{app:pxk}
    \mathcal{P}(\tilde{x},\tilde{\kappa})=\left.-\frac{1}{\beta N}\frac{\delta }{\delta g(\tilde{x},\tilde{\kappa})} \langle \log(\mathcal{Z})\rangle\right|_{g=0}.
\end{equation}

\subsection{The replica trick and the replicated partition function}\label{App:replica}

\subsubsection{The replicated partition function}
To be able to make use of Eq.~\eqref{app:pxk} we need to obtain an expression for the logarithm of the partition function. We use the replica trick
\begin{equation}\label{app:replica_trick}
    \langle \log(\mathcal{Z})\rangle=\left.\frac{d}{d n} \langle \mathcal{Z}^n\rangle \right|_{n=0}
\end{equation}
where $\mathcal{Z}^n$ represents the $n$-fold replicated partition function
\begin{equation}
    \langle \mathcal{Z}^n\rangle = \int  \Pi_{i} \left( d\kappa_i dh_i P(\kappa_i)P(h_i) \right) \int \Pi_{i}\Pi_{\alpha}dx^\alpha_i   e^{-\beta \sum_\alpha \mathcal{H}^S[x^\alpha]}
\end{equation}
and $\mathcal{H}^S[x^\alpha]$ represents the Hamiltonian of one of the replicas (we use Greek letters to label different copies, $\alpha\!=\!1, 2, \dots, n$). We first massage the expression of the replicated partition function into something more manageable: with two consecutive Hubbard--Stratonovich transformations we remove the coupling between oscillators and we perform the averaging over the random fields ${h_i}$, obtaining the following expression
\begin{eqnarray}
    \langle \mathcal{Z}^n\rangle&=&\left(\frac{N}{2\pi} \right)^{n/2} \int  \, (\Pi_\alpha dz_\alpha) e^{- \frac{N}{2}\sum_\alpha z_\alpha^2 +N \log( \mathcal{Z}_1)}
\end{eqnarray}
with
\begin{equation}
    \mathcal{Z}_1=\int  \, (\Pi_\alpha dx^\alpha)d\kappa \, ds \, \frac{e^{-s^2/2}}{\sqrt{2\pi}}P(\kappa)     e^{\sum_\alpha \left[  z_\alpha \sqrt{\beta J}x^\alpha-\beta  \left(\frac{A}{4!} (x^\alpha)^4-h x^\alpha + g(x^\alpha,\kappa) + \kappa  \frac{1}{2}(x^\alpha)^2 - s \sigma  x^\alpha\right) \right]}
\end{equation}
Note that it was not necessary to introduce any overlap matrix, and that the different replicas do not interact with each other. The fields $s$ and $z_\alpha$ come from the Hubbard--Stratonovich transformations. Note also that we now introduced the auxiliary field $g(x^\alpha, \kappa)$ in the Hamiltonian from the beginning.

\subsubsection{The replica trick in the thermodynamic limit}

In the thermodynamic limit we can apply the saddle-point approximation, and the expression of the replicated partition function simplifies to
\begin{equation}\label{eq:saddle_point_approx}
    \langle \mathcal{Z}^n\rangle \sim e^{- N S[z^*_\alpha]} \quad ,
\end{equation}
where the replicated action reads
\begin{equation}
    S[z_\alpha]=\frac{1}{2}\sum_\alpha z_\alpha^2-\log( \mathcal{Z}_1)
\end{equation}
and $z^*_\alpha$ corresponds to the minimum of the $S[\bullet]$ function. We omit the proportionality factor in Eq.~\eqref{eq:saddle_point_approx}, which would just be the inverse of the square root of the determinant of the Hessian (with respect to $z_\alpha$) of the replicated action, and does not play a role here.

The value of $z_\alpha^*$ that satisfies the saddle-point equation $\frac{\partial S[z_\alpha^*]}{\partial z_\gamma}\!=\!0$ is given by the implicit equation
\begin{equation}\label{app:eq_for_z}
    z =\frac{ \sqrt{\beta J}}{\mathcal{Z}_1} \int  d\kappa \, ds \, \frac{e^{-s^2/2}}{\sqrt{2\pi}}P(\kappa)   \left(\int dx\,  e^{  -\beta V_{\rm eff}(x)} \right)^{n-1}\left(\int dx\, x e^{-\beta V_{\rm eff}(x)}\right) 
\end{equation}
where by symmetry we imposed that $z^*_\alpha\!=\!z$ for all $\alpha$, and where we defined the effective potential
\begin{equation}
    V_{\rm eff} (x)=  \frac{A}{4!} x^4  + \frac{\kappa}{2}x^2 - \left(s  \sigma + h + z \sqrt{\frac{J}{\beta}} \right)x + g(x, \kappa)
\end{equation}
The replicated partition function is then given by 
$\langle \mathcal{Z}^n\rangle\!\sim\!e^{- N S[z]}$, where $S[z]\!=\!\frac{ n}{2} z^2\!-\!\log( \mathcal{Z}_1[z])$, and
\begin{equation}
    \mathcal{Z}_1 [z]=\int  d\kappa \, ds \, \frac{e^{-s^2/2}}{\sqrt{2\pi}} P(\kappa)   \left(\int dx \, e^{  -\beta V_{\rm eff}(x)} \right)^n 
\end{equation}
Now, we can apply the replica trick, given by Eq.~\eqref{app:replica_trick}
\begin{equation}
    \langle \log(\mathcal{Z})\rangle=\left.\frac{d}{d n} \langle \mathcal{Z}^n\rangle \right|_{n=0}= N \left(- \frac{1}{2} z^2 - n z \frac{d z}{d n} + \frac{1}{ \mathcal{Z}_1[z]} \frac{d}{d n}  \mathcal{Z}_1[z]\right)\left.e^{- N S[z]}\right|_{n=0} \quad .
\end{equation}
It is possible to show that all the terms in the sum on the right-hand side are well defined in the $n\!\to\!0$ limit. In particular, Eq.~\eqref{app:eq_for_z} provides a finite value for the first element of the sum, $z_0\!=\!\lim_{n\to 0} z$
\begin{equation}\label{app:eq_for_z0}
    z_0 = \sqrt{\beta J}\int d\kappa \, ds \,  \frac{e^{-s^2/2}}{\sqrt{2\pi}}P(\kappa)   \frac{\int dx \, x e^{  -\beta  V_{\rm eff}(x)}  }{\int dx \,  e^{ -\beta V_{\rm eff}(x)} } \quad .
\end{equation}
The second term, instead, vanishes since the first derivative of $z$ with respect to $n$ is finite when evaluated in zero. Moreover, we have that $\left.\mathcal{Z}_1[z]\right|_{n=0}\!=\!1 $ and $\left.S[z]\right|_{n=0}\!=\!0$. Therefore we only need to study
\begin{equation}
    \frac{d}{d n}  \mathcal{Z}_1[z]= \int d\kappa \, ds \, \frac{e^{-s^2/2}}{\sqrt{2\pi}}P(\kappa)   \left(\int dx \, e^{ -\beta V_{\rm eff}(x)}  \right)^n \left[ \log \left(\int dx \, e^{-\beta   V_{\rm eff}(x)} \right)+n \frac{d}{d n} \log \left(\int dx \,  e^{ -\beta V_{\rm eff}(x)} \right) \right]
\end{equation}
Again, the last term of the expression can be shown to be null in the limit $n\!\to\!0$ since the derivative stays finite, therefore we obtain
\begin{equation}\label{app:eq:logZ}
 \langle \log(\mathcal{Z})\rangle= N \left(- \frac{1}{2} z_0^2 +  \int d\kappa \, ds \, \frac{e^{-s^2/2}}{\sqrt{2\pi}}P(\kappa)  \log\left( \int dx\,  e^{ -\beta V_{\rm eff}(x)}  \right) \right)   
\end{equation}
We can, then, obtain the joint probability $\mathcal{P}(\tilde{x},\tilde{\kappa})$ using expression \eqref{app:pxk}, which yields
\begin{equation}\label{app:eq_P_X_K}
    \mathcal{P}(\tilde{x},\tilde{\kappa}) = P(\tilde{\kappa})\int  ds \, \frac{e^{-s^2/2}}{\sqrt{2\pi}} \frac{  e^{ - \beta V_{\rm eff}(\tilde{x},\tilde{\kappa})}  }{ \int dx \, e^{ - \beta V_{\rm eff}(x,\tilde{\kappa})}}
\end{equation}

Finally, we can perform the following rescaling $z_0\!\to\!\sqrt{\beta J} z_0$, in order to remove the explicit $\beta$ dependence in the expression of the effective potential, which now reads
\begin{equation}\label{app:eq_eff_pot_resc}
    V_{\rm eff} (x)=  \frac{A}{4!} x^4  + \frac{\kappa}{2}x^2 - \left(s  \sigma + h + J z_0 \right)x
\end{equation}
Equations \eqref{app:eq_for_z0} and \eqref{app:eq:logZ} are also (minimally) affected by this rescaling. We spell them out for completeness: these are the equations used in the rest of the study and that we presented in Sec.~\ref{simplified_khgps} of this paper.
\begin{align}
    \langle \log(\mathcal{Z})\rangle & = N \left(- \frac{\beta J}{2} z_0^2 +  \int d\kappa \, ds \, \frac{e^{-s^2/2}}{\sqrt{2\pi}}P(\kappa)  \log\left( \int dx\,  e^{ -\beta V_{\rm eff}(x)}  \right) \right)   \\
    z_0 &= \int d\kappa \, ds \,  \frac{e^{-s^2/2}}{\sqrt{2\pi}}P(\kappa)   \frac{\int dx \, x e^{  -\beta  V_{\rm eff}(x)}  }{\int dx \,  e^{ -\beta V_{\rm eff}(x)} } \quad .\label{app:eq_for_z0_rescaled}
\end{align}

\subsubsection{The low-temperature limit}

We proceed with taking the low-temperature limit, $\beta\!\to\!\infty$, of the expressions \eqref{app:eq_P_X_K} and \eqref{app:eq_for_z0_rescaled}. This allows us to compute the integrals over the variable $x$ through a saddle-point approximation. The self-consistency equation now reads
\begin{equation}\label{eq_zeroTemp_z2}
    z_0= \int d\kappa \,P(\kappa)\int ds \, \frac{e^{-s^2/2}}{\sqrt{2\pi}}   x_0
\end{equation}
where $x_0\!=\!x_0 (\kappa, s ; z_0)$ is the global minimum of the effective potential, $\frac{d V_{\rm eff}(x,\kappa ; z_0)}{dx}|_{x=x_0}\!=\!0$. The equation for the distribution is, instead, given by
\begin{equation}\label{app:eq:p_x_k}
    \mathcal{P}(\tilde{x},\tilde{\kappa})=P(\tilde{\kappa})\int ds \,   \frac{e^{-s^2/2}}{\sqrt{2\pi}} \delta(\tilde{x} - x_0)
\end{equation}
We can make use of the composition property of the delta function
\begin{equation}\label{app:delta_property}
    \delta(f(s)) = \sum_i \frac{\delta(s- s_i)}{|f'(s_i)|}
\end{equation}
with $f(s)\!=\!\tilde{x}\!-\!x_0(s;\tilde{\kappa}, z_0)$ and where the prime denotes a derivative with respect to $s$. We just need to find the value (or values) of $s_i$ such that, for a given $\tilde{x}$ and $\tilde{\kappa}$, we have $x_0 (s_i ; \tilde{\kappa}, z_0)\!=\!\tilde{x}$. The expression of the effective potential gives us the following relation between the minimum $x_0$ and $s$
\begin{equation}
    \left. \frac{d V_{\rm eff}}{dx} \right| _{x=x_0} = \frac{A}{3!} x_0^3 + \kappa x_0 - (h + J z_0 + s \sigma) = 0
\end{equation}
which we can readily invert in favor of $s$, obtaining a single value $s_0$ which produces that minimum $x_0\!=\!\tilde{x}$:
\begin{equation}\label{eq:s_0}
    s_0(x_0=\tilde{x})=\frac{1}{\sigma}\left( \frac{A}{3!} \tilde{x}^3 + \tilde{\kappa} \tilde{x} - ( J z_0 + h) \right) \quad .
\end{equation}
 We are finally able to compute the integral \eqref{app:eq:p_x_k}
\begin{equation}\label{app:eq:p_x_k_2}
    \mathcal{P}(\tilde{x},\tilde{\kappa})= P(\tilde{\kappa}) \frac{e^{-s_0^2/2}}{\sqrt{2\pi}|f'(s_0)|}
\end{equation}

Before concluding this section we want to comment on the support of $\mathcal{P}$. The question is whether there exists a value $s_0$ for any given pair $(\tilde{x},\tilde{\kappa})$. This is the case for a positive value of $\kappa$, since the minimum $x_0$ changes continuously over the whole real axis. On the other hand, the potential has two minima when $\kappa$ is negative, and in this case the global minimum $x_0$ jumps discontinuously (see Fig.~\ref{fig:minima_s}). It is possible to obtain the edges of the skipped interval on the $\tilde{x}$ axis by setting the coefficient of the first order term of the potential to zero, since there is no cubic term in Eq.~\eqref{app:eq_eff_pot_resc}. The support of $\mathcal{P}$ reads
\begin{equation}
    \Supp(\mathcal{P}) = \left(\tilde{x} \in \mathbb{R} \, \land \,\tilde{\kappa} \geq 0 \right) \lor \left[ \left (\tilde{x} < -  \sqrt{- 6 \tilde{\kappa}/A} \, \lor \, \tilde{x} >  \sqrt{- 6 \tilde{\kappa}/A}  \right) \land (\tilde{\kappa}< 0)\right] \quad .
\end{equation}

%%%%%%%%%%%%%%%%%%%%%%%%%%%
\begin{figure}
    \centering
    \includegraphics[width=0.8\linewidth]{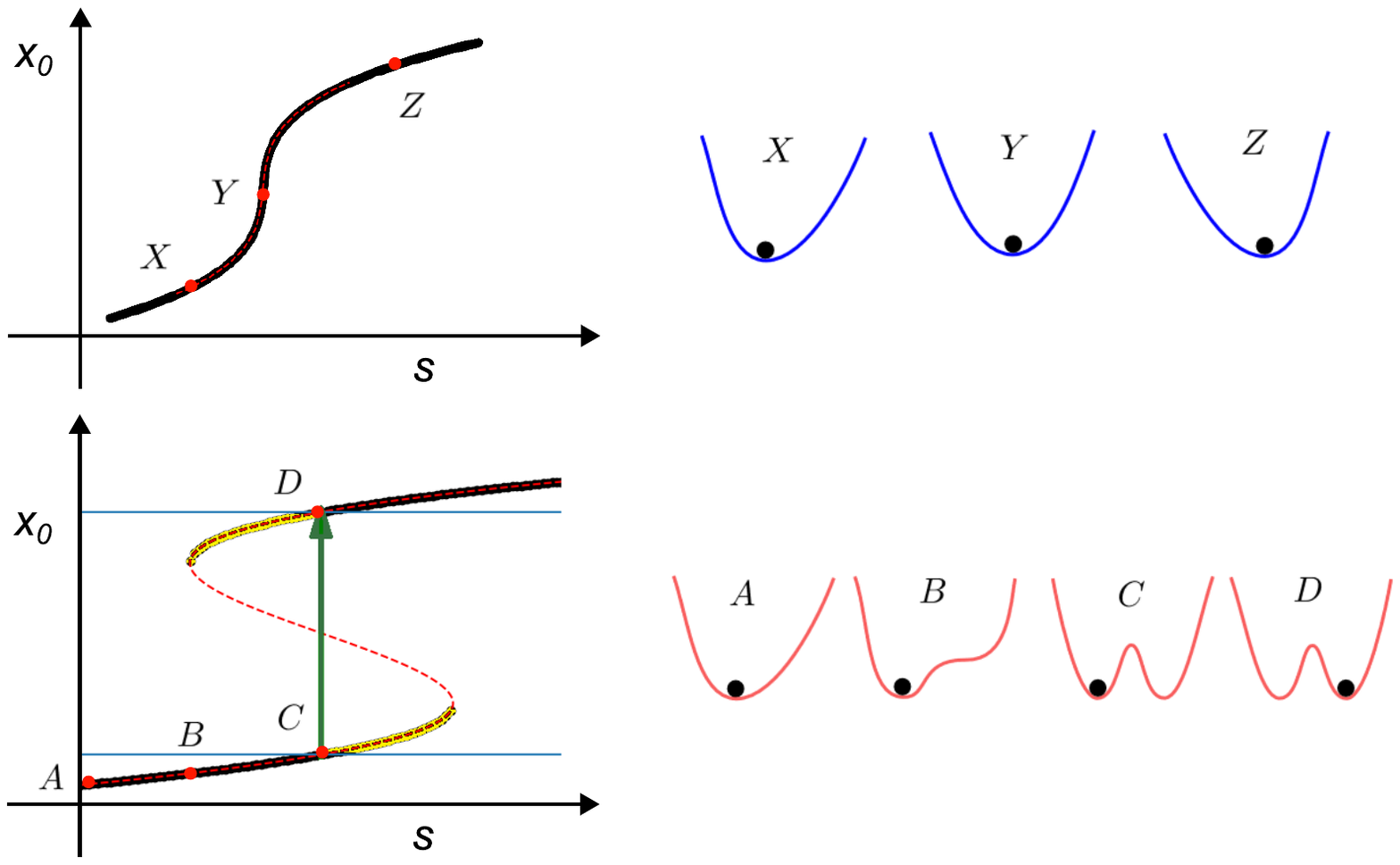}
    \caption{(Above) For a positive support of the stiffness distribution ($\kappa_{min}\!\geq\!0$), the minimum $x_0$ (shown on the plot on the left as a function of $s$) can assume any real value, since the effective potential (on the right) has a single minimum: varying $s$ just moves the minimum around, in a continuous way. (Below) When the support of the stiffness distribution extends to the negative part of the real axis, the potential develops a second minimum. For negative $\kappa$, at a certain value of $s$ the position of the global minimum will \textit{discontinuously} jump from the location of the first local minimum to the other one. This jump in the function $x_0 (s)$ corresponds to a jump in the domain of $s_0(x_0)$. Hence there are values of $x$ that the minimum $x_0$ can never assume (for certain values of $\kappa$). For these values the probability $\mathcal{P}(x,\kappa)$ is identically zero.}
    \label{fig:minima_s}
\end{figure}
%%%%%%%%%%%%%%%%%%%%%%%%%%%

\subsection{The density of states for the simplified KHGPS}\label{app:Dos_var}

With the expression of $\mathcal{P}(x, \kappa)$ in our hands we can compute the distribution of eigenvalues of the diagonal matrix. Plugging Eq.~\eqref{app:eq:p_x_k_2} in the expression for $\mathcal{P}\left(\sqrt{\frac{2}{A}(\lambda - \kappa)}, \kappa\right)$ in the spectrum, Eq.~\eqref{app:spectrum_expr}, we obtain
\begin{equation}\label{dos_intermediate}
    \rho_\mathcal{M}(\lambda) = \frac{1}{\sqrt{2 \pi}}\int_{\max(-\frac{\lambda}{2}, \kappa_{m})}^{\min(\lambda,\kappa_{M})} d\kappa \, \frac{P(\kappa)}{\sqrt{2A(\lambda - \kappa)}} \left(  \frac{e^{-(s_{0}^+)^2/2}}{|f'(s_{0}^+)|}  + \frac{e^{-(s_{0}^-)^2/2}}{|f'(s_{0}^-)|} \right)
\end{equation}
where we now dropped the tildes to ease the notation, and we defined
\begin{equation}\label{app:eq_s0_pm}
    s_{0}^\pm \equiv s_0 (\pm x, \kappa) =\frac{1}{\sigma} \left[ -J z_0 - h \pm (A x^3 / 6 + \kappa x) \right] \quad .
\end{equation}
Note that we also used the information on the support of $\mathcal{P}$ to bound the integration interval from below: the condition $ x^2\!>\!(-6\kappa/A)$, together with the fact that $x\!=\!\sqrt{\frac{2}{A}(\lambda - \kappa)}$, translates to $\kappa\!>\!-\lambda/2$ .

We can further simplify the expression by evaluating the denominator $|f'(s_0)|$. The derivative of the minimum $x_0$ with respect to $s$ can be obtained from the inverse relation, Eq.~\eqref{eq:s_0}, which yields the following simple expression for both $s_0^{\pm}$
\begin{equation}
    |f'(s_0)| = \left.\frac{d x_0}{ds} \right|_{s=s_0} = \frac{\sigma}{\frac{A}{2}x^2 + \kappa } = \frac{\sigma}{\lambda} \quad .
\end{equation}
Performing the change of variable $\kappa\!\to\!\lambda \kappa$, the density of eigenvalues of the Hessian reduces to
\begin{equation}\label{dos_step2}
    \rho_\mathcal{M}(\lambda) = \frac{\lambda^{3/2}}{2 \sigma\sqrt{ A \pi}}\int_{\max(-\frac{1}{2}, \kappa_{m}/\lambda)}^{\min(1,\kappa_{M}/\lambda)} d\kappa \, \frac{P(\lambda \kappa)}{\sqrt{1 - \kappa}} \left( e^{-(s_{0}^+)^2/2}  + e^{-(s_{0}^-)^2/2} \right)
\end{equation}
where after the rescaling of the variable $\kappa$, it is useful to redefine $x$:
\begin{equation}
    x = \sqrt{\frac{2}{A}(\lambda - \lambda \kappa)} = \sqrt{\lambda} \sqrt{\frac{2}{A}(1-\kappa)} \equiv \sqrt{\lambda} \, u \quad .
\end{equation}
Note that with this redefinition $u$ does not depend on the eigenvalue $\lambda$. Expression \eqref{app:eq_s0_pm} now reads
\begin{equation}
    s_{0}^\pm  =\frac{1}{\sigma} \left[ -J z_0 - h \pm \lambda^{3/2} (A u^3 / 6 + \kappa u) \right] \quad .
\end{equation}

Next, we need to specify the stiffness distribution $P(\kappa)$. We choose a generic power-law distribution with exponent $\alpha$. Properly normalized, it reads
\begin{equation}\label{app:eq:stiff_distr}
    P(\kappa) = \frac{\alpha + 1}{(\kappa_{M} - \kappa_{m})^{\alpha + 1}} (\kappa - \kappa_{m})^\alpha
\end{equation}
Note that when $\kappa_m\!<\!0$ the distribution has a finite value in zero: $P(0)\!>\!0$. The eigenvalue spectrum becomes
\begin{equation}
    \rho_\mathcal{M}(\lambda) =\frac{\lambda^{3/2}}{2 \sigma\sqrt{A \pi}}  \frac{\alpha + 1}{(\kappa_{M} - \kappa_{m})^{\alpha + 1}} \int_{\max(- \frac{1}{2}, \kappa_{m}/\lambda)}^{\min(1,\kappa_{M}/\lambda)} d\kappa \, \frac{(\lambda\kappa - \kappa_{m})^\alpha}{\sqrt{1 - \kappa}} \left(  e^{-(s_{0}^+)^2/2}  + e^{-(s_{0}^-)^2/2} \right)  
\end{equation}
From the support of the distribution and from the integration domain, we notice that $\rho_\mathcal{M}(\lambda)$ is gapped for $\kappa_m\!>\!0$, and that the gap closes when $\kappa_m\!=\!0$. We focus now on the region $\kappa_m\!<\!0$, which shows the most interesting behavior. We are interested in the low-energy spectrum of excitations. In this limit, $\lambda\!\ll\!1$, we can safely ignore the exponential dependence on $\lambda$ inside the integral, so that
\begin{equation}
    \exp\left({-\frac{(s_{0}^\pm)^2}{2}}\right) \approx \exp\left({-\frac{(J z_0 + h)^2}{2\sigma^2}}\right)
\end{equation}
The binomial term $(\lambda \kappa\!-\!\kappa_m)^\alpha$ needs to be carefully considered.  In the limit of $\lambda\!\to\!0$, this term is just $(-\kappa_{m})^\alpha$ (which is positive since we are limiting ourselves to the case $\kappa_m\!<\!0$) and the density of eigenvalues can be approximated by
\begin{equation}
     \rho_\mathcal{M}(\lambda) \approx \frac{\lambda^{3/2}}{\sigma} (-\kappa_{m})^\alpha e^{-(Jz_0+h)^2/(2\sigma^2)} I_0(\alpha) \quad ,
\end{equation}
with
\begin{equation}
    I_0(\alpha) =  \frac{1}{\sqrt{A \pi}} \frac{\alpha + 1}{(\kappa_{M} - \kappa_{m})^{\alpha + 1}} \int_{- \frac{1}{2}}^{1} d\kappa \, \frac{1}{\sqrt{1 - \kappa}} =  \frac{\sqrt{6}}{\sqrt{A \pi}} \frac{\alpha + 1}{(\kappa_{M} - \kappa_{m})^{\alpha + 1}}  \quad .  
\end{equation}
When $\lambda$ grows, the contribution given by $\lambda \kappa$ will increase until it takes over the whole binomial term, which will be dominated by $(\lambda \kappa)^\alpha$. In this regime we have
\begin{equation}
     \rho_\mathcal{M}(\lambda) \approx \frac{\lambda^{3/2 + \alpha}}{\sigma} e^{-(Jz_0+h)^2/(2\sigma^2)} I_1(\alpha)
\end{equation}
with
\begin{equation}
    I_1(\alpha) =  \frac{1}{ \sqrt{A \pi}} \frac{\alpha + 1}{(\kappa_{M} - \kappa_{m})^{\alpha + 1}} \int_{0}^{1} d\kappa\, \frac{\kappa^\alpha}{\sqrt{1 - \kappa}}  =  \frac{B\left(\alpha+1,\frac{1}{2}\right)}{ \sqrt{A \pi}} \frac{\alpha + 1}{(\kappa_{M} - \kappa_{m})^{\alpha + 1}}  \quad ,
\end{equation}
where $B(\alpha\!+\!1, \frac{1}{2})$ is the Beta function. Finally, in terms of the density of states, $\mathcal{D}(\omega)\!=\!2 \omega \rho_\mathcal{M}(\omega^2)$, we obtain that at low frequencies the system is characterized by a quartic scaling of 
\begin{equation}
    \mathcal{D}(\omega) \approx  \omega^{4} \frac{2 I_0(\alpha)(-\kappa_{m})^\alpha}{\sigma}e^{-\frac{(h+Jz_0)^2}{2\sigma^2}}  \quad ,
\end{equation}
while at higher frequencies it crosses over to
\begin{equation}
    \mathcal{D}(\omega) \approx  \omega^{4+2\alpha} \frac{2 I_1(\alpha)}{\sigma}e^{-\frac{(h+Jz_0)^2}{2\sigma^2}}  \quad .
\end{equation}
The crossover frequency is given by
\begin{equation}
    \omega_{\text{cross}} = \sqrt{-\kappa_{m}} \left( \frac{I_0(\alpha)}{I_1(\alpha)} \right)^{\frac{1}{2\alpha}}
\end{equation}
which to leading order scales like $\omega_{\text{cross}}\!\sim\!\sqrt{-\kappa_m}$.

\end{widetext}

\bibliography{references}

\end{document}